\documentclass[]{aastex6}

\newcommand*\mean[1]{\overline{#1}}
\newcommand{\MS}{\ifmmode{\,}\else\thinspace\fi{\rm M}\ifmmode_{\odot}\else$_{\odot}$\fi}
\newcommand{\LS}{\ifmmode{\,}\else\thinspace\fi{\rm L}\ifmmode_{\odot}\else$_{\odot}$\fi}
\newcommand{\RS}{\ifmmode{\,}\else\thinspace\fi{\rm R}\ifmmode_{\odot}\else$_{\odot}$\fi}

\begin{document}


\title{Mass and p-factor of the type II Cepheid OGLE-LMC-{\allowbreak}T2CEP-098 in a binary system
\footnote{This paper includes data gathered with the 6.5m
Magellan Clay Telescope at Las Campanas Observatory, Chile.}
}

\author{Bogumi{\l} Pilecki\altaffilmark{1,2}}
\affil{Copernicus Astronomical Centre, Polish Academy of Sciences, Bartycka 18, 00-716 Warsaw, Poland}

\author{Wolfgang Gieren\altaffilmark{4}}

\affil{Universidad de Concepci{\'o}n, Departamento de Astronom{\'i}a,Casilla 160-C, Concepci{\'o}n, Chile}

\author{Rados{\l}aw Smolec}
\affil{Copernicus Astronomical Centre, Polish Academy of Sciences, Bartycka 18, 00-716 Warsaw, Poland}

\author{Grzegorz Pietrzy{\'n}ski\altaffilmark{3}}
\affil{Copernicus Astronomical Centre, Polish Academy of Sciences, Bartycka 18, 00-716 Warsaw, Poland}

\author{Ian B. Thompson}
\affil{Carnegie Observatories, 813 Santa Barbara Street, Pasadena, CA 91101-1292, USA}

\author{Richard I. Anderson}
\affil{Physics and Astronomy Department, The Johns Hopkins University, 3400 North Charles St, Baltimore, MD 21202, USA}

\author{Giuseppe Bono\altaffilmark{5}}
\affil{Dipartimento di Fisica Universit`a di Roma Tor Vergata, viadella Ricerca Scientifica 1, 00133 Rome, Italy}

\author{Igor Soszy{\'n}ski}
\affil{Warsaw University Observatory, Al. Ujazdowskie 4, PL-00-478, Warszawa, Poland}

\author{Pierre Kervella\altaffilmark{6}}
\affil{Unidad Mixta Internacional Franco-Chilena de Astronomía (CNRS UMI 3386), Departamento de Astronomía, Universidad de Chile, Camino El Observatorio 1515, Las Condes, Santiago, Chile}

\author{Nicolas Nardetto}
\affil{Laboratoire Lagrange, UMR7293, Universit{\'e} de Nice Sophia-Antipolis, CNRS, Observatoire de la C{\^o}te d’Azur, Nice, France}

\author{M{\'o}nica Taormina\altaffilmark{2}}
\affil{Copernicus Astronomical Centre, Polish Academy of Sciences, Bartycka 18, 00-716 Warsaw, Poland}

\author{Kazimierz St\c{e}pie{\'n}}
\affil{Warsaw University Observatory, Al. Ujazdowskie 4, PL-00-478, Warszawa, Poland}

\author{Piotr Wielg{\'o}rski}
\affil{Copernicus Astronomical Centre, Polish Academy of Sciences, Bartycka 18, 00-716 Warsaw, Poland}

\altaffiltext{1}{pilecki@camk.edu.pl}
\altaffiltext{2}{Instituto de Astronomía Te{\'o}rica y Experimental, (IATE, CONICET-UNC), C{\'o}rdoba, Argentina}
\altaffiltext{3}{Universidad de Concepci{\'o}n, Departamento de Astronom{\'i}a,Casilla 160-C, Concepci{\'o}n, Chile}
\altaffiltext{4}{Millenium Institute of Astrophysics, Santiago, Chile}
\altaffiltext{5}{INAF - Osservatorio Astronomico di Roma, Via Frascati 33, 00040 Monte Porzio Catone, Italy}
\altaffiltext{6}{LESIA (UMR 8109), Observatoire de Paris, PSL Research University, CNRS, UPMC, Univ. Paris-Diderot, 5 Place Jules Janssen, 92195 Meudon, France}

\begin{abstract}

We present the results of a study of the type II Cepheid ($P_{puls}=4.974$ d) in the eclipsing binary system OGLE-LMC-T2CEP-098 ($P_{orb}=397.2$ d). The Cepheid belongs to the peculiar W Vir group, for which the evolutionary status is virtually unknown. It is the first single-lined system with a pulsating component analyzed using the method developed by Pilecki et al. (2013).  We show that the presence of a pulsator makes it possible to derive accurate physical parameters of the stars even if radial velocities can be measured for only one of the components.
We have used four different methods to limit and estimate the physical parameters, eventually obtaining precise results by combining pulsation theory with the spectroscopic and photometric solutions. The Cepheid radius, mass and temperature are $25.3\pm{}0.2$ $R_\odot$, $1.51\pm{}0.09$ $M_\odot$ and $5300\pm{}100$ K, respectively, while its companion has similar size ($26.3 R_\odot$), but is more massive ($6.8$ $M_\odot$) and hotter ($9500$ K). Our best estimate for the p-factor of the Cepheid is $1.30\pm{}0.03$.
The mass, position on the period-luminosity diagram, and pulsation amplitude indicate that the pulsating component is very similar to the Anomalous Cepheids, although it has a much longer period and is redder in color. The very unusual combination of the components suggest that the system has passed through a mass transfer phase in its evolution. More complicated internal structure would then explain its peculiarity.

\end{abstract}



\section{Introduction}
\label{sec:intro}

Type II Cepheids are low-mass pulsating stars that belong to the disc and halo populations \citep{2002PASP..114..689W}. They are a much older counterpart of the more massive classical Cepheids -- they have periods and amplitudes in a similar range, but are about 1.5-2 mag fainter. They exhibit a tight and well-defined period-luminosity relation and may serve as a good distance indicator, allowing a measurement of distances both inside and outside of our Galaxy (\citealt{2011MNRAS.413..223M}, \citealt{2008AcA....58..293S}, \citealt{2009AcA....59..403M}).

Compared to classical Cepheids, our knowledge of type II Cepheids is very poor, being more qualitative than quantitative. Type~II Cepheids are usually divided into three subgroups depending on the pulsation period, observational properties and evolutionary status, but nevertheless present a similar period-luminosity relation.
Starting from the work of \citet{1976ApJ...204..116G,1985MmSAI..56..169G} it is generally accepted that those with the shortest periods (called BL Herculis / BL Her stars) are evolving from the Horizontal Branch to the Asymptotic Giant Branch (AGB). Those with periods in the range of 4-20 days (called W Virginis / W Vir stars) are on their way up the AGB and enter the pulsational instability strip due to helium shell flashes, which makes them move to higher temperatures on the HR diagram. Finally, those with the longest periods (called RV Tauri / RV Tau stars) are leaving the AGB on their way to the white dwarf cooling sequence. The result of Gingold's work was only qualitative though, explaining the difference between these three subgroups with different periods, without an explanation of the relative rate of occurrence nor a measurement of the values of basic parameters including the
masses of these variables.
The situation is even more complicated as more recent evolutionary models \citep[see][and references therein]{2016CoKon.105..149B} do not predict thermal pulses that would explain the existence of W Vir stars.

It is clearly important to obtain direct measurements of the masses for a sample of type II Cepheids to pinpoint their evolutionary status. The best means for such measurements are eclipsing binary systems in which one or both components are pulsating stars. We have applied this method to eclipsing binaries containing classical Cepheids obtaining very precise masses (\citealt{cep227mnras2013, cep2532apj2015}; \citealt{cep9009apj2015}).

The same eclipsing binary method also enables a direct determination of another important parameter, namely the projection factor (hereafter p-factor), which is the conversion factor between the observed radial velocity and the stellar pulsational velocity. This methodology has been used to measure the p-factor for three classical Cepheids (\citealt{cep227mnras2013}; \citealt{cep9009apj2015}; Pilecki et al. in prep).

As part of research on the characteristics of type II Cepheids the OGLE project \citet{2008AcA....58..293S} identified a group of W Virginis stars that for similar periods have differently looking light curves, with the rising branch being steeper than the declining one. These were called peculiar W Virginis stars (hereafter also pWVir). In general these stars also lie above the normal sequence for type II Cepheids and are bluer in color. For a statistically significant fraction eclipses and ellipsoidal modulations were detected. The authors suggested that all of these stars are members of binary systems.

OGLE-LMC-T2CEP-098 (hereafter T2CEP-098) is one of the peculiar W Virginis stars that exhibit eclipses. Its existence was first reported by \citet{1999IAUS..190..513W}. The system was later analyzed by \citet{macho2002alcock}, but due to the lack of spectroscopic data and good modeling tools only approximate results were obtained. They noted that the components have similar radii, but that the companion to the Cepheid is much brighter and bluer. They could not say anything about the absolute physical parameters however.

In this study we present high-resolution spectra and use more sophisticated modeling together with pulsation theory to obtain a consistent picture of the system and the stars of which it is composed, including the important physical parameters (like mass) of the Cepheid T2CEP-098.

\section{Data}
\label{sec:data}

\begin{deluxetable}{lccccc}
\tablecaption{Basic data of the OGLE-LMC-T2CEP-098 system.}
\tablewidth{0pt}
\tablehead{
\colhead{Parameter} & \colhead{Value} & \colhead{Unit} & \colhead{Source}
}
\startdata
Right Ascension     &  05:20:25.0	 & hh:mm:ss.s & OGLE-3 \\
Declination         & -70:11:08.7    & dd:mm:ss.s & OGLE-3 \\
$K_{S,vmc}$         & 13.89          & mag     & VMC    \\
$I_{C,ogle}$        & 14.37          & mag     & OGLE-3 \\
$R_{C,macho}$       & 14.59          & mag     & MACHO  \\
$V_{ogle}$          & 14.67          & mag     & OGLE-3 \\
$V_{macho}$         & 14.71          & mag     & MACHO  \\
$P_{orb}$           & 397.2          & days    & OGLE-3 \& MACHO \\
$P_{puls}$          & 4.974          & days    & OGLE-3 \& MACHO \\
\enddata
\label{tab:basic}
\tablecomments{ Source references: VMC \citep{2015MNRAS.446.3034R}, OGLE-3 \citep{2008AcA_Udalski_OGLEIII}, MACHO \citep{1999PASP..111.1539A}.}
\end{deluxetable}

\begin{figure}
\begin{center}
  \resizebox{0.6\linewidth}{!}{\includegraphics{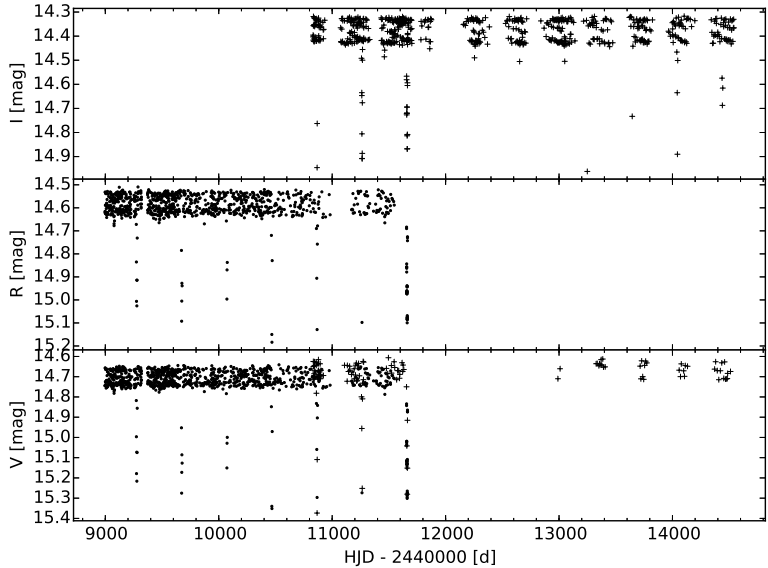}} \\
\caption{All photometric data collected for OGLE-LMC-T2CEP-098. Data from the OGLE project are marked with crosses and data from the MACHO project with dots.
\label{fig:obs_all}}
\end{center}
\vspace{0.5cm}
\end{figure}

Our analysis makes use of 678 measurements in the I-band and 121 in the V-band acquired with the Warsaw telescope at Las Campanas Observatory by the OGLE project (\citealt{2003AcA....53..291U}, \citealt{2008AcA....58..293S}). We have also used instrumental V-band (940 points) and R-band (953 points) data from the MACHO project \citep{1999PASP..111.1539A} downloaded from their webpage\footnote{http://macho.anu.edu.au} and converted to the Johnson-Cousins system using equations from \cite{2007AJ....134.1963F}. Basic data for the system are given in Table~\ref{tab:basic} , while in Fig.~\ref{fig:obs_all} we present all of the photometric data used in the modeling.

In contrast to our recent analysis of OGLE LMC562.05.9009 \citep{cep9009apj2015} we did not transform the MACHO $V$-band light curve to the OGLE system because the OGLE $V$-band data are quite scarce and cannot serve as a reference. A reverse transformation (to the MACHO system) is not advisable as the MACHO data, apart from being quite noisy, are taken in non-standard filters.  As a result we have modeled only the OGLE $I$ and MACHO $R$ and $V$ data, and use the OGLE $V$-band light curve to obtain the flux ratio of the component stars, once the other parameters were determined.

The K-band data for this system come from the VISTA
near-infrared $YJK_S$ survey of the Magellanic Clouds \citep[VMC,][]{2011_VMC_Cioni}. No eclipses were observed and so the light curve cannot be used for modeling. The K-band data were used to assist in the calculation of the effective temperature of the component stars. 

The spectroscopic data were acquired using the MIKE spectrograph on the 6.5-m Magellan Clay telescope at Las Campanas Observatory in Chile, the HARPS spectrograph attached to the 3.6-m telescope at La Silla Observatory and the UVES spectrograph on VLT at Paranal Observatory. In total we obtained 30 high-resolution spectra (21 MIKE + 6 HARPS + 3 UVES). The MIKE data were reduced using Daniel Kelson's pipeline available at the Carnegie Observatories Software Repository\footnote{http://code.obs.carnegiescience.edu/}. The  UVES data were reduced using ESO Reflex software and the official pipeline available at the ESO Science Software repository\footnote{http://www.eso.org/sci/software.html}. For HARPS we have used the online reduced data which are  available at the ESO Archive\footnote{http://archive.eso.org/}.

\added{All the used light curves and radial velocity measurements are available online\footnote{http://users.camk.edu.pl/pilecki/p/t2c098}.}

\section{Initial analysis}
\label{sec:init}

\subsection{Photometry}
\label{sub:photo}

\begin{figure}
\begin{center}
  \resizebox{0.55\linewidth}{!}{\includegraphics{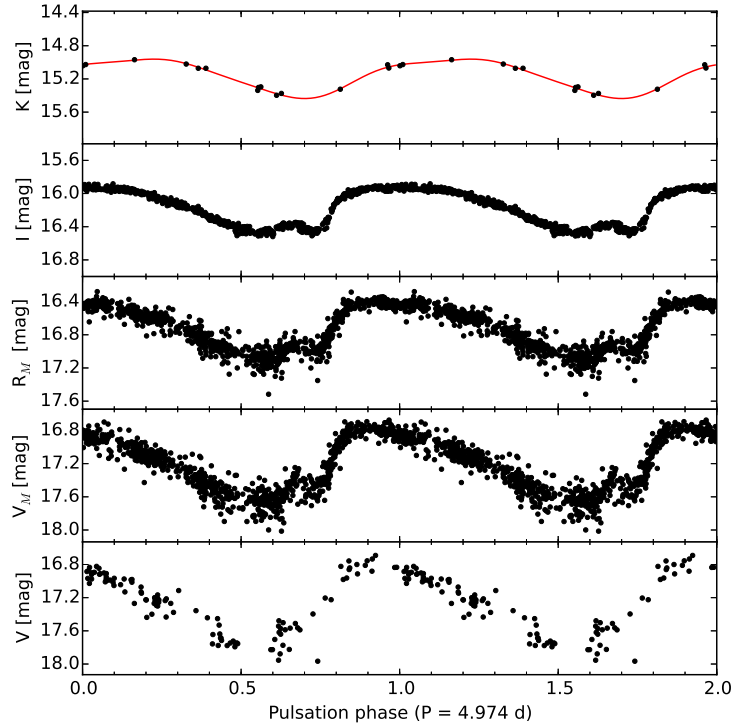}} \\
\caption{Pulsational light curves of the Cepheid OGLE-LMC-T2CEP-098 with the light of the companion subtracted, folded with the ephemeris $T_{max} ($HJD$) = 2455500.00 + 4.973726\times $E. Y-axis span is the same (1.6 mag) for all the panels. For K-band the 3-order Fourier fit is shown as well.
\label{fig:lcpuls}}
\end{center}
\end{figure}

The pulsational period was determined using the OGLE I-band  and MACHO V-band  light curves together after removing the phases that correspond to eclipses. We did not use the MACHO R-band data as the measurements are taken at the same times and do not bring much  new information. A cyclical period change was also detected corresponding to the orbital motion of the Cepheid and this was taken into account in the fit. The light curves were also detrended in the process, i.e. we have subtracted the systematic brightness deviations that are not related to the pulsational or eclipsing variability. In most cases these were seasonal variations with a period of about one year.

The pulsational period obtained from the MACHO data is a bit lower than the one obtained from the OGLE-III data with the difference being larger than expected for the estimated uncertainties. In the analysis we have used a weighted mean value of both, $P_{mean} = 4.973726 d$.

Out-of-eclipse light curves of T2CEP-098 folded with the pulsation period are plotted in Fig.~\ref{fig:lcpuls}. The shape of the $I$-band light curve of the Cepheid was analyzed through its Fourier decomposition parameters \citep{1981ApJ...248..291S}. To describe the light curve shape the amplitude ratios $R_{i1} = a_i/a_1$ and phase differences $\phi_{i1} = \phi_i - i*\phi_1$ are used, with a series of the following form fitted to the data:

$$ a_0 + \sum\limits_{i=1}^{n} a_i \cos\left[i\omega(t-t_0) + \phi_i\right] $$

\begin{figure}
\begin{center}
  \resizebox{0.5\linewidth}{!}{\includegraphics{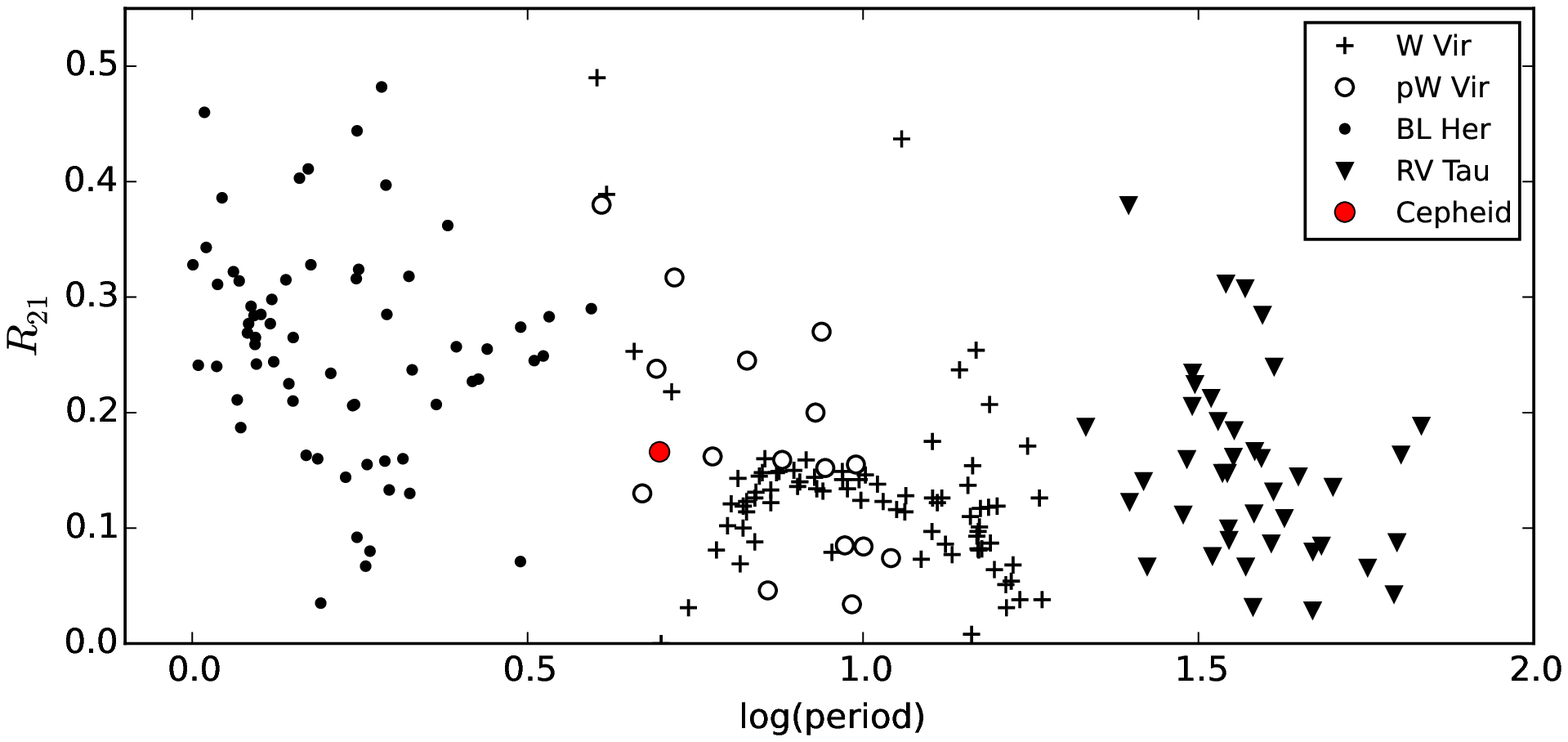}} \\
  \resizebox{0.5\linewidth}{!}{\includegraphics{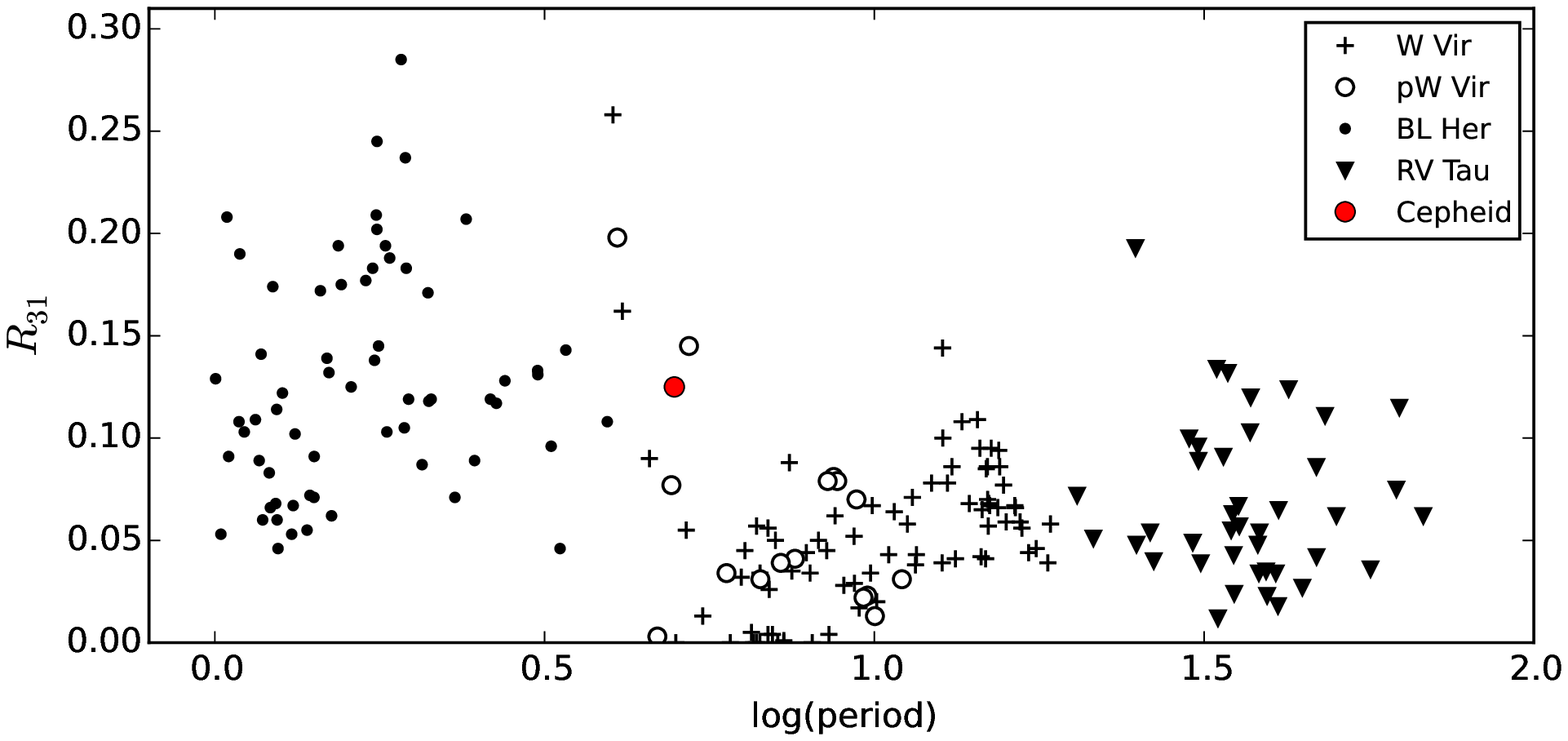}} \\
\caption{Fourier decomposition -- $R_{21}$, $R_{31}$ parameters. OGLE-LMC-T2CEP-098 is marked by a red-filled circle.
\label{fig:four_r}}
\end{center}
\end{figure}

\begin{figure}
\begin{center}
  \resizebox{0.5\linewidth}{!}{\includegraphics{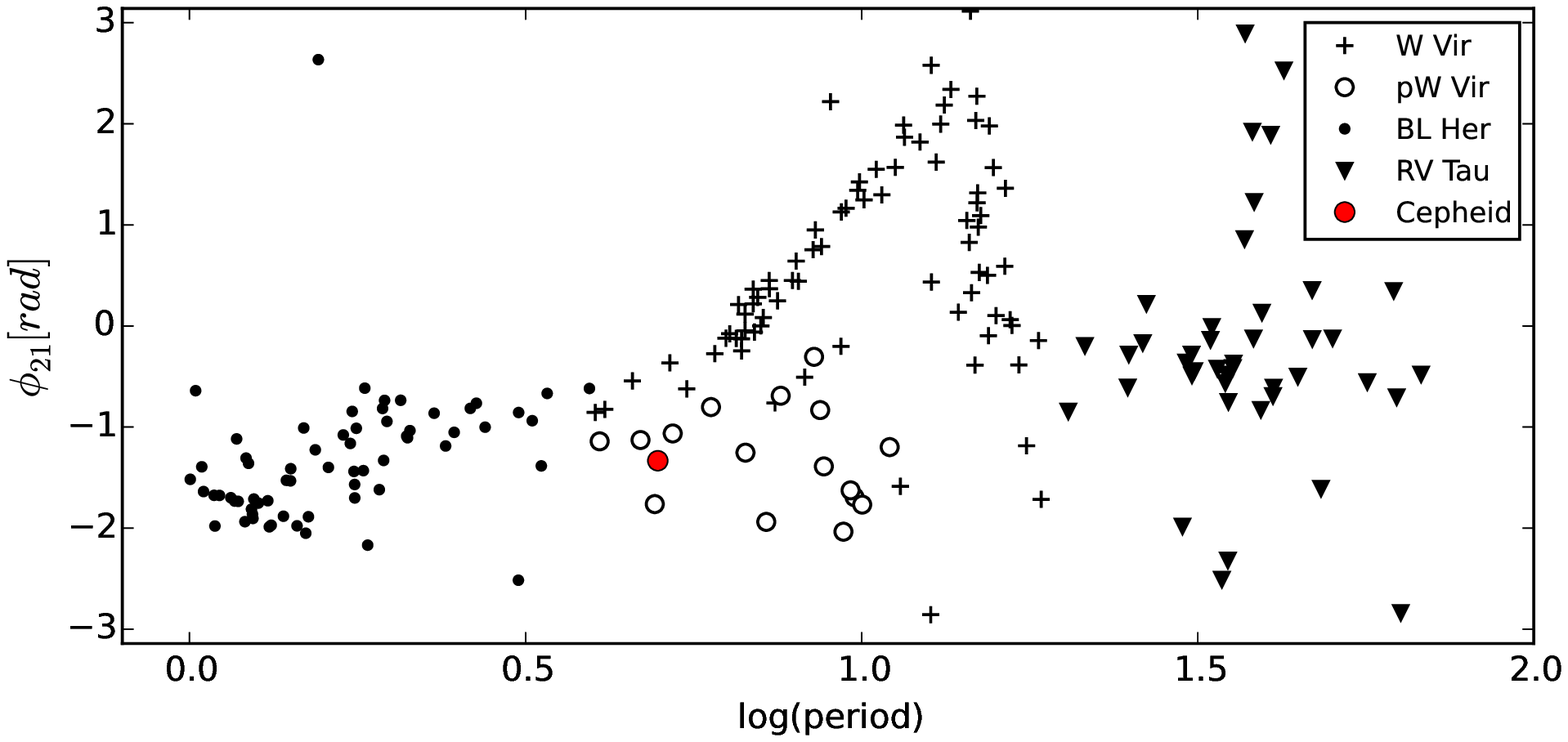}} \\
  \resizebox{0.5\linewidth}{!}{\includegraphics{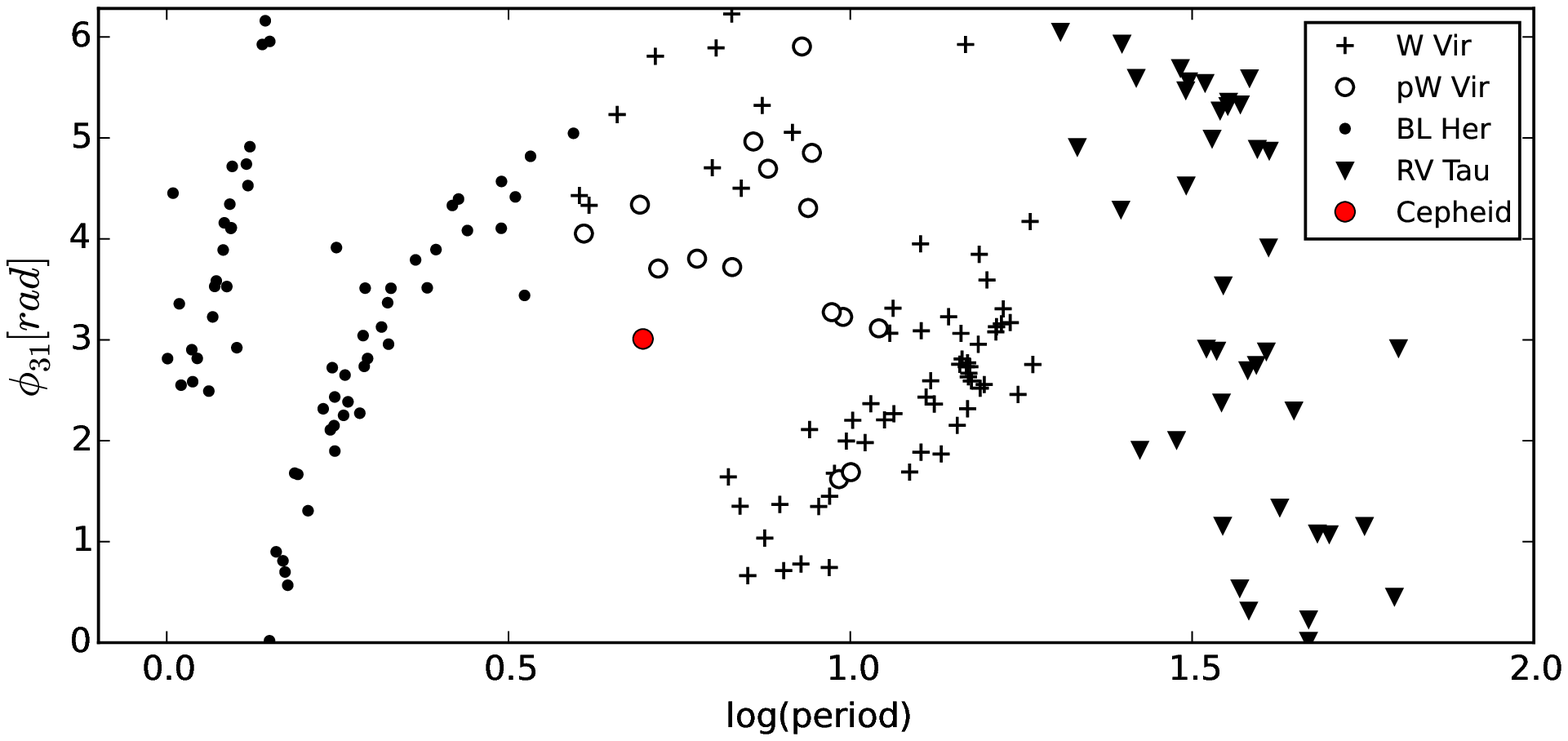}} \\
\caption{Fourier decomposition -- $\phi_{21}$ and $\phi_{31}$ parameters. OGLE-LMC-T2CEP-098 is marked by a red-filled circle.
\label{fig:four_phi}}
\end{center}
\end{figure}

The position of the star in the $R_{21}$, $R_{31}$, $\phi_{21}$ and $\phi_{31}$ vs. period plane is compared to other Type II Cepheids in Fig.~\ref{fig:four_r} and \ref{fig:four_phi}. The pulsational period corresponds to the range occupied by W Virginis stars, but as  noted by \cite{2008AcA....58..293S}, the light curve shape is different compared to that of a typical star of this type with a similar period. As a result, these authors classified the T2-CEP-098 as a peculiar W Vir star.

However, the light curve shape (compared with the light curves in Fig.~3 of \citealt{2008AcA....58..293S}) and the position of T2CEP-098 on the diagrams suggest that it may have more in common with BL Herculis type stars. BL Herculis stars  with periods $P \sim 2d$ have almost identical light curves\footnote{see also: \texttt{http://ogle.astrouw.edu.pl/atlas/BL\_Her.html}} compared to T2CEP-098.
\cite{2008AcA....58..293S} add that the distinction between the BL Her and W Vir stars is not clear and different authors suggest different limiting periods ranging from 3 to 10 days \citep[e.g][]{2016JAVSO..44..179N}.

The situation is further complicated since the light curve of the pulsating component is also similar to those of classical Cepheids with $P \sim 17d$. This observation together with other features of the star revealed by the analysis presented below make its classification even more difficult. We will elaborate more on this in Section~\ref{sub:class}.

\begin{figure}
\begin{center}
  \resizebox{0.6\linewidth}{!}{\includegraphics{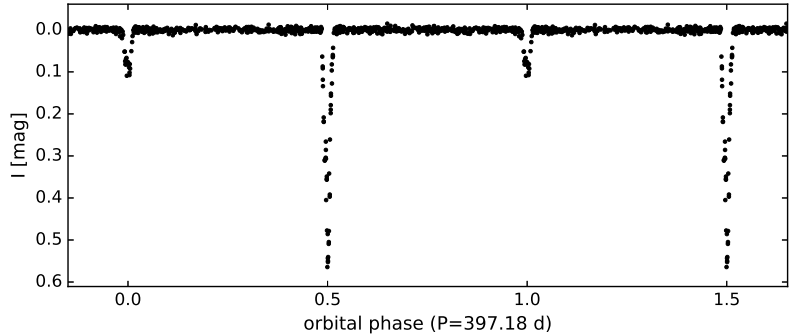}} \\
\caption{The I-band orbital light curve of the system, freed from the pulsations of the Cepheid. Both eclipses are well covered. The large difference in eclipse depths indicates a large difference in the temperatures of the component stars.
\label{fig:lcorb}}
\end{center}
\end{figure}

Once the pulsational light curves were prepared they were subtracted from the raw data to obtain the eclipsing light curves. The I-band light curve prepared this way (presented in Fig.~\ref{fig:lcorb}) was used to obtain an initial photometric solution in order to have a rough estimation of the most important parameters. \added{This estimation was then used as a starting point in the main analysis.}

\pagebreak

\subsection{Radial velocities}
\label{sub:radvel}

T2CEP-098 is a single-lined spectroscopic binary (SB1) and only velocities of the Cepheid could be extracted from the spectra. \added{Although the presence of the companion was detected in the Balmer lines, velocities measured from these lines were not accurate enough to derive its orbit. However they scatter about the systemic velocity of the system and are clearly anti-correlated with the orbital velocity of the Cepheid.}

In several spectra other sources were detected at constant velocities: about 252 km/s, 288 km/s and 325 km/s. These additional sources are barely visible in the blue part of the spectrum, but are clearly evident in the red portion of the spectrum. To increase the detection efficiency \added{of the additional components} we have searched for them in the wavelength range 5350 to 6800 $\mathring{A}$, narrower and redder than the region used to determine radial velocities (RVs) of the Cepheid.

\begin{figure}
\begin{center}
  \resizebox{0.5\linewidth}{!}{\includegraphics{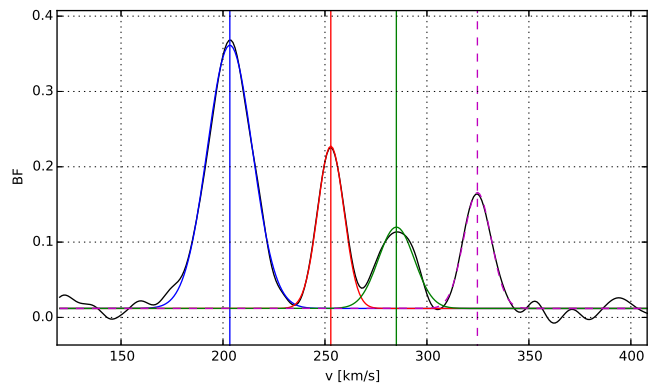}} \\
\caption{Broadening Function profiles of T2CEP-098 (blue) and other sources detected in one of the spectra in the wavelength range 5350 to 6800 $\mathring{A}$. Note that the radial velocities of the Cepheid were measured in a much broader range 4125 to 6800 $\mathring{A}$, for which these additional sources were barely visible. When detected, the additional peaks were included in the profile fit at known fixed velocities, making their influence negligible.
\label{fig:extlight}}
\end{center}
\end{figure}

\begin{figure}
\begin{center}
  \resizebox{0.3\linewidth}{!}{\includegraphics{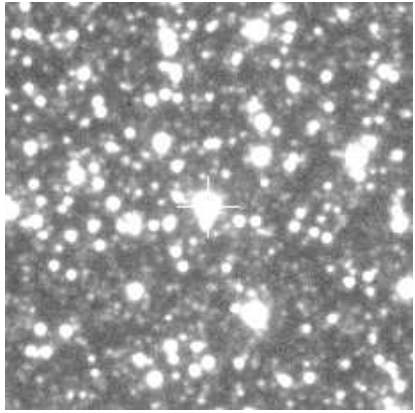}} \\
\caption{Neighborhood of OGLE-LMC-T2CEP-098 (marked with a white cross). The field is very crowded, which enhances the possibility for nearby stars to enter the slit and thus appear in the spectra. An area of 60x60 arcseconds is shown.
\label{fig:stars}}
\end{center}
\end{figure}

For one of the spectra the profiles of the additional sources obtained this way are even comparable to the Cepheid profile (see Fig.~\ref{fig:extlight}). 
However prominent third light is excluded by the analysis of the light curves described in Section~\ref{sec:analysis}. As the system is located in a very dense field (see Fig.~\ref{fig:stars}) it is highly probable that light from nearby stars on the sky has entered the slit, thus affecting only the obtained spectra and not the photometry.
\added{Note that the slit for the MIKE spectrograph cannot be set since the instrument is fixed with respect to the Nasmyth platform of the Magellan Clay telescope. As a result the position angle of the slit on the sky rotates during any individual observation.}

\begin{deluxetable}{lr@{ $\pm$ }lc}
\tablecaption{Orbital solution for OGLE-LMC-T2CEP-098.}
\tablewidth{0pt}
\tablehead{
\colhead{Parameter} & \multicolumn{2}{c}{Value}  & \colhead{Unit}
}
\startdata
$T_0-2450000$   & 5830.7391$^a$ &     -    &  days   \\
orbital period  &  397.1781$^a$ &     -    &  days   \\
$\gamma$        &  263.5$^b$    &    0.4   &  km/s   \\ 
$K_1$           &   47.7        &    0.4   &  km/s \\ 
$e$             &    0.0$^a$    &  (0.015) &  - \\
$a_1 \sin i$    &    375        &     3    &  $R_\odot$ \\ 
$m_2 \sin^3 i$  & $>$4.47       &    0.10  &  $M_\odot$ \\
pulsation period&    4.97327    &  0.00005 &  days \\
rms$_1$         & \multicolumn{2}{c}{0.91} &  km/s 
\label{tab:spec}
\enddata
\tablecomments{$T_0$ and period were taken from the photometric solution. Eccentricity $e$ was set to 0, consistent with the both photometric and orbital solutions. The eccentricity error is taken into account in estimating errors in the other parameters.}
\tablenotetext{a}{ fixed value }
\tablenotetext{b}{ at the given epoch (changes due to the third body influence) }

\end{deluxetable}

To measure velocities we used the Broadening Function method \citep{1992AJ....104.1968R, 1999ASPC..185...82R} implemented in the RaveSpan application \citep{cep227mnras2013, cep2532apj2015} with a template matching the Cepheid in the temperature-gravity plane. The template was taken from the library of theoretical spectra of \citet{2005A&A...443..735C}. Radial velocities were measured in the range 4125 to 6800 $\mathring{A}$. 
In this wavelength range the Cepheid is the only strong, variable peak in the majority of spectra. In some cases, where additional light sources were also visible we have included them in the profile fitting with fixed velocities. The overall effect of these sources on the measured Cepheid velocities is thus negligible. The typical formal errors of the Cepheid velocities derived this way are $\sim 900$ m/s.

The measured velocities show cyclical variations with two periods corresponding to the pulsational and orbital motions. This unambiguously confirms that the Cepheid T2CEP-098 is a component of an eclipsing binary system.

To disentangle the orbital and pulsational motions we simultaneously fitted the eccentricity $e$, argument of periastron $\omega$, velocity semi-amplitude $K_1$, systemic velocity $\gamma$, and a number of Fourier series coefficients representing the pulsational radial velocity (RV) curve . The orbital period $P_{orb}$ and the time of the spectroscopic conjunction $T_0$ was kept fixed at the values taken from the preliminary analysis of the photometry. After some trials, the order of the Fourier series was set to $n=4$, which gave the best fit without overfitting.
Higher orders might  be necessary if the star happens to have a more complicated RV curve, but this cannot be confirmed with the present data.

After fitting this model a systematic difference in the residuals was seen (with higher velocities at the beginning of the observations and lower at the end), suggesting the influence of a third body with a period of the order of ten thousand days. Assuming an external period ($P_{ext}$) of 15000 days and zero eccentricity we then added the $K_{12}$ amplitude of the system center-of-mass to the model and obtained a value of $5 \pm 1$ km/s and a much better fit to the data with no systematic difference in the residuals. The amplitude is highly correlated with the assumed period, but this assumption has no impact on the other parameters.

All of the aforementioned additional objects detected in the spectra showed constant velocity within the errors, so it is highly unlikely that one of them is the supposed third component of the system, as it would have to be much more massive than the stars in the eclipsing binary together. 

\begin{figure}
\begin{center}
  \resizebox{0.6\linewidth}{!}{\includegraphics{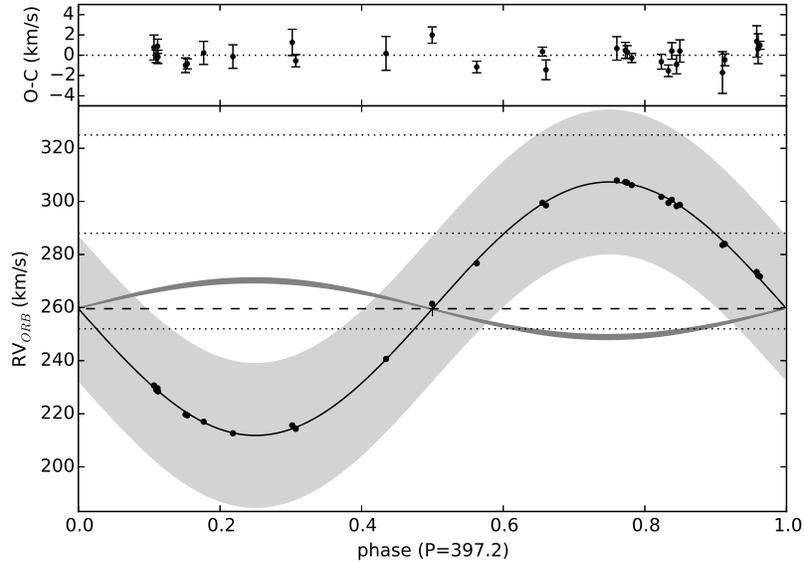}} \\
\caption{Measured radial velocities of the Cepheid (points) with the pulsations removed overplotted on the orbital solution. The pulsational variability range of the Cepheid is marked with light gray area, while the expected RV for the companion velocity is marked with dark gray area. Detected additional sources with constant velocities are shown as dotted lines.
\label{fig:rvorb}}
\end{center}
\end{figure}


\begin{figure}
\begin{center}
  \resizebox{0.6\linewidth}{!}{\includegraphics{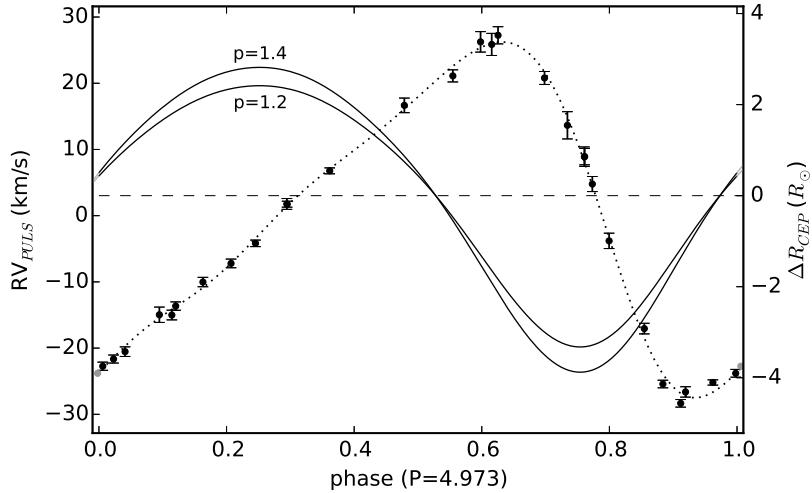}} \\
\caption{Pulsational radial velocity curve (points) and the radius variation (solid lines) of the Cepheid over one pulsation cycle. To obtain this RV curve the orbital motion was removed from the measured radial velocities. The full range of $\Delta R$ is $5.7-6.7 R_\odot$ depending on the assumed p-factor (1.2-1.4). 
\label{fig:rvpuls}}
\end{center}
\end{figure}

The results were used to obtain the final photometric model. In the process the orbital solution was updated using new values for $P_{orb}$ and $T_0$ for full consistency. The eccentricity was also fixed at 0.0 as obtained from the photometry (and consistent with the orbital solution within errors). The final model parameters are presented in Table~\ref{tab:spec} and the orbital and pulsational radial velocity curves are shown in Fig.~\ref{fig:rvorb} and \ref{fig:rvpuls}, respectively.  The high pulsational radial velocity amplitude (about 50 km/s) translates to a radius change of $5.7-6.7 R_\odot$ depending on the assumed p-factor (1.2-1.4).

\subsection{Pulsational period}
\label{sub:pulsper}

Erratic period changes are known to be present among type II Cepheids \citep{2016JAVSO..44..179N}. They were noted for example for RU Cam \citep{1998PASP..110.1428P}, $\kappa$ Pav (\citealt{2008MNRAS.386.2115F}, \citealt{2015A&A...576A..64B}) and various Cepheids in globular clusters (\citealt{2010AJ....139.2300R}, \citealt{2017MNRAS.465..173P}). W Virginis itself also exhibits nonlinear period changes, but in this case the changes are quite regular and are probably an effect of multiperiodicity \citep{2007AJ....134.1999T}.
On the other hand, our object is not a typical type II Cepheid, and so we must be cautious in taking these examples as a guide.

\begin{figure}
\begin{center}
  \resizebox{0.6\linewidth}{!}{\includegraphics{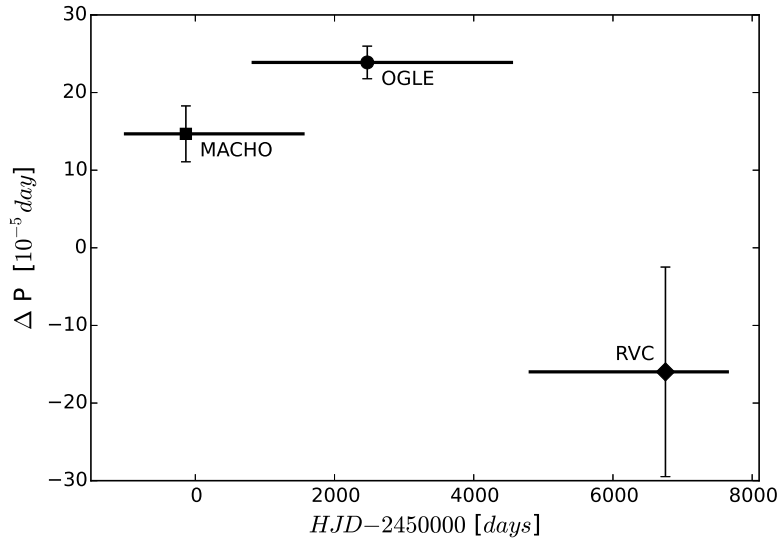}} \\
\caption{Periods measured for different data sets ($\Delta P = P - 4.9735078$ days). Horizontal lines shows the timespan of the given data set, the point is located at the weighted center of the data set and the measured period (with errors). RVC marks the period measured from the radial velocity curve.
\label{fig:dper}}
\end{center}
\end{figure}

The initial analysis of the  photometric data indeed led to the discovery of different pulsational periods for all analyzed datasets (see Fig.~\ref{fig:dper}) inconsistent with a linear period change. As already mentioned, we have detected a small but significant difference in the periods obtained from the analysis of MACHO and OGLE light curves. Although the difference cannot be explained by the errors of the period determination a linear period change could explain it.

Much more striking however is the low value of the pulsation period obtained from analysis of the radial velocities. This is only marginally consistent (within 3 sigma) with the pulsational periods obtained from the light curves and it does not follow the expected slow period increase.
On the other hand no significant linear nor long-term cyclical period change was detected in any of the data sets. We have checked the possibility that for some reason the period from RVs is wrong. The most probable explanation is an unfortunate phase distribution of our measurements.

If we adopt a fixed period obtained from the photometry, the fit of the model to the measured radial velocities is not satisfactory unless we significantly increase (from 4 to 7-8) the order of the Fourier series that describes the pulsations. Since the number of RV measurements is not high enough, some overfitting occurs. The shape of the resulting pulsational RV curve is a bit different with a small bump close to the maximum (roughly corresponding in phase with the bump seen in the light curves).  The resulting difference in the radius change and the Cepheid orbital amplitude is not significant. We conclude that there are no effects of using a fitted (lower) pulsational period on the final results.
At the moment the results obtained on the period change of the Cepheid are not conclusive and a longer observational baseline would be needed to confirm the erratic period change.

\section{Analysis and results} \label{sec:analysis}

\subsection{Analysis} \label{sub:analysis}

For a detailed description of the method and the analysis steps we refer the reader to \cite{cep227mnras2013} and the papers on other Cepheids that follow. Here we will present a short summary of the subject.

The photometric data were analyzed using a pulsation-enabled eclipsing modeling tool based on the well-tested {\tt JKTEBOP} code \citep{jktebop2004southworth}, modified to allow the inclusion of pulsational variability. The {\tt JKTEBOP} code itself is based on the EBOP code \citep{1981AJ.....86..102P}, in which the stars are treated as spheres for calculating eclipse shapes and as bi-axial ellipsoids for calculating proximity effects. This treatment of the stars makes it useful only for well separated components, but the code works very fast and has extremely low numerical noise. 

In our approach we generate a two-dimensional light curve that consists of purely eclipsing light-curves for different pulsating phases. Then a one-dimensional light curve is generated (interpolated from the grid) using a combination of pulsational and orbital phases calculated with the Cepheid and system ephemerides.

From the photometric solution we have the period, the time of the primary minimum ($T_{I}$), the inclination ($i$), the fractional radii ($r_1$ and $r_2$ calculated from the fitted sum and ratio of radii), the eccentricity ($e$), the argument of periastron ($\omega$), the surface brightness ratios ($j_{21}$), the third light ($l_3$), and the p-factor ($p$) value. Fractional radii are expressed in the units of $A$, the separation of the components, such that $r_i = R_i/A$, where $R_i$ are the physical radii of the components.

\begin{figure}
\begin{center}
  \resizebox{0.9\linewidth}{!}{\includegraphics{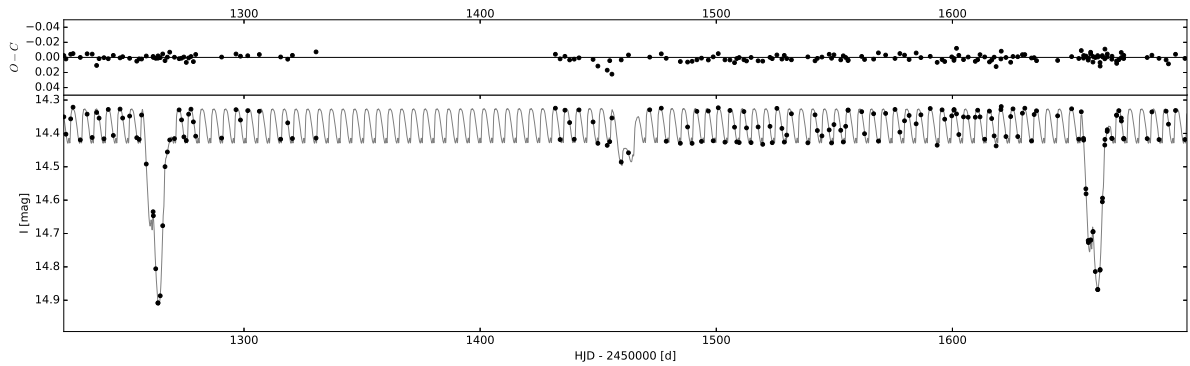}} \\
\caption{OGLE I-band light curve model covering two eclipses of a companion ({\it deep ones}) and one eclipse of the Cepheid ({\it shallow one}).
\label{fig:oglei_mod}}
\end{center}
\end{figure}

\begin{figure}
\begin{center}
  \resizebox{0.48\linewidth}{!}{\includegraphics{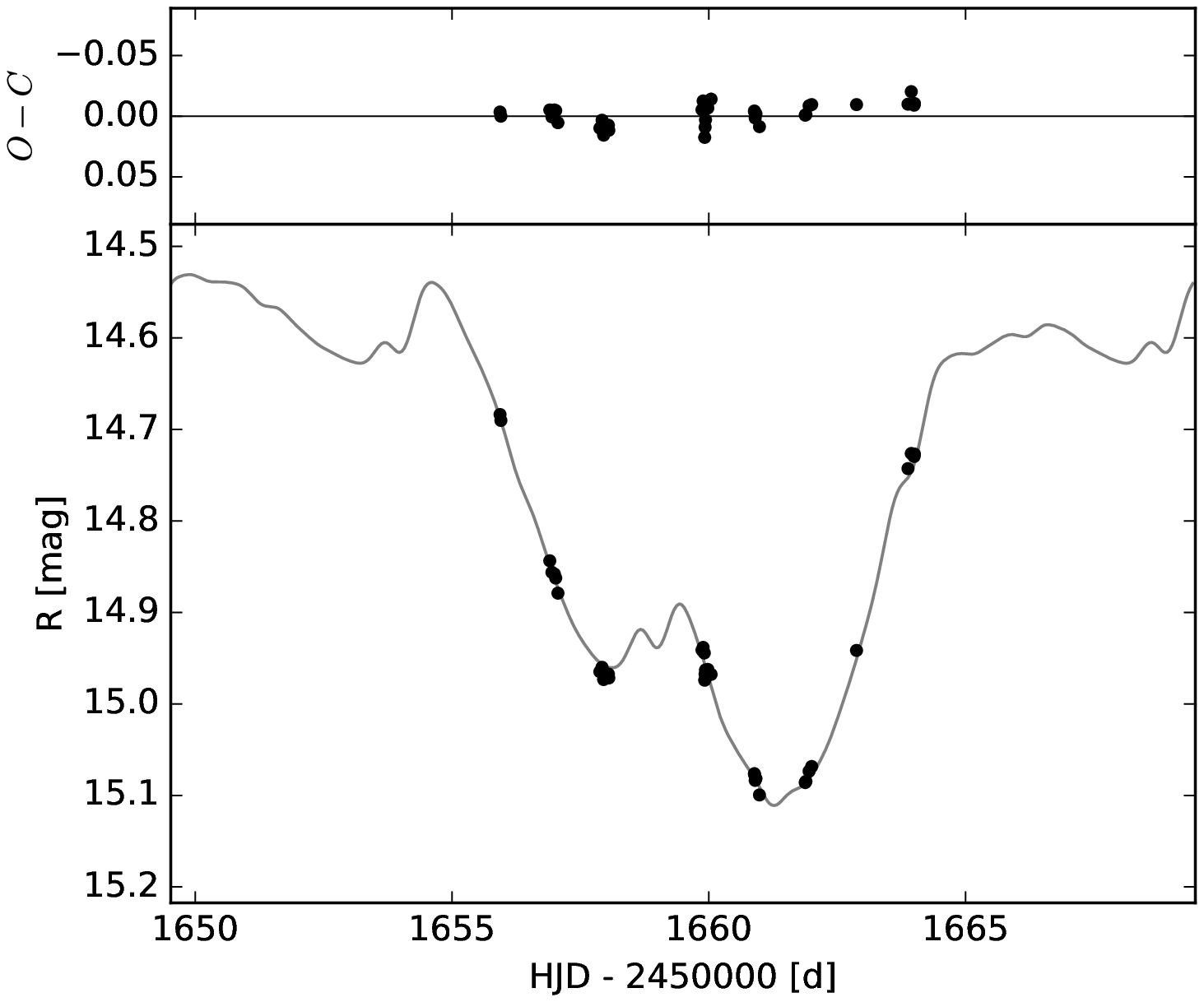}} 
  \resizebox{0.48\linewidth}{!}{\includegraphics{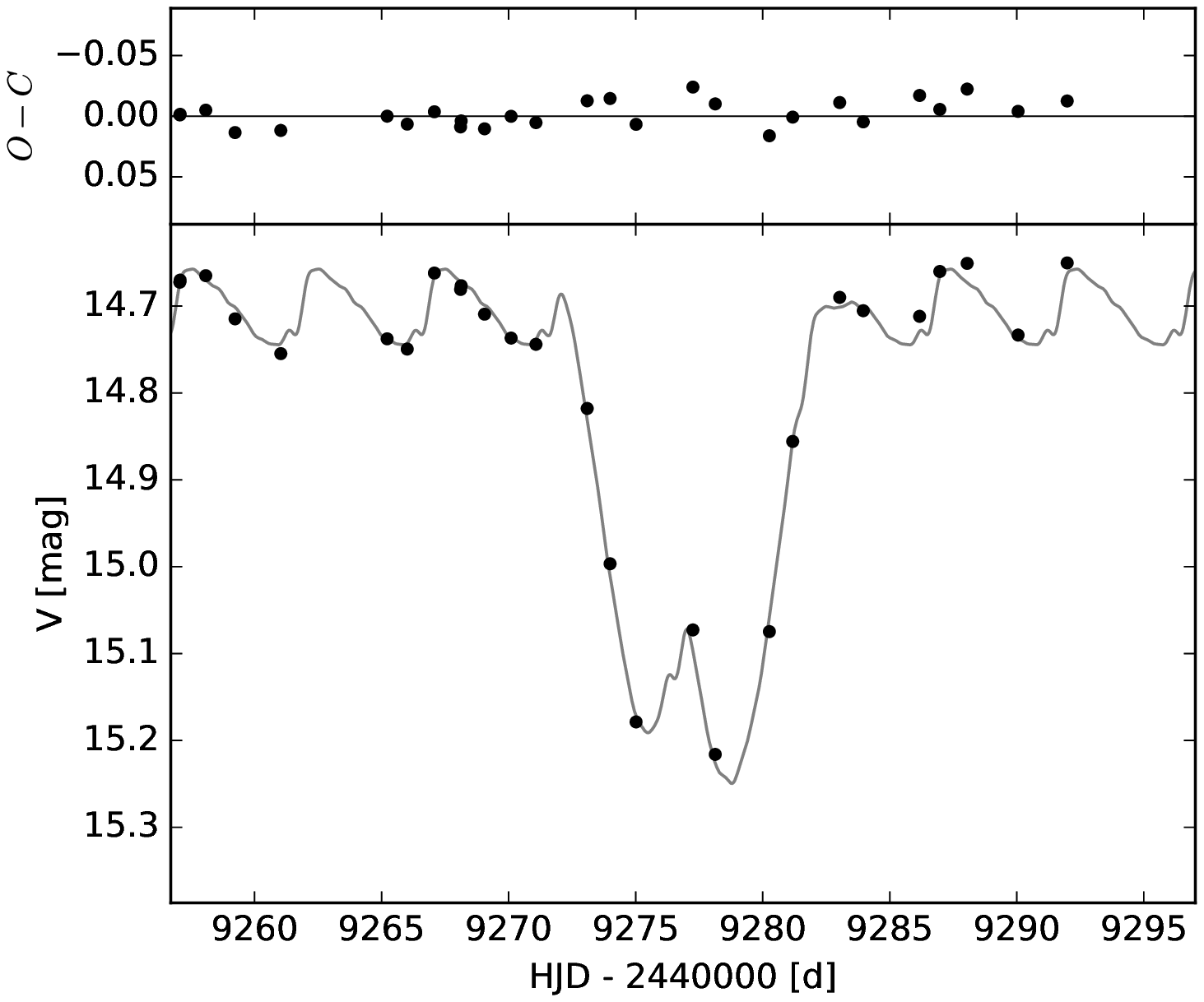}} \\
\caption{MACHO R-band ({\it left}) and V-band ({\it right}) light curve models centered on selected eclipses.
\label{fig:machor_ecl}}
\end{center}
\end{figure}

In the {\tt JKTEBOP} code limb darkening is described by a set of atmospheric parameters, including the temperature, which can be set fixed or fitted.
At the beginning we set the temperatures to 5000 and 8500 as obtained from the initial solution and the observed color-temperature transformation. Once the final solution was reached we fixed all of the parameters and solved for the temperatures that give the limb darkening parameters that best describe the data. In this way we obtained  a temperature of $5500_{-400}^{+200}$ K for the Cepheid and $9500_{-500}^{+1000}$ K for the companion.

In our model we can only calculate the p-factor with respect to the known separation ($A$) of the stars in the system. Because in this case we only know the size of the orbit of the Cepheid ($A_1 = 375 R_\odot$) we had to assume a value for $A$ in order to do the analysis. A separation of 437 $R_\odot$ was chosen as it corresponds to a Cepheid mass of 1~$M_\odot$ and a p-factor of 1.24 -- a reasonable compromise between the expected values for these two parameters. The expected mass for type II Cepheids is ~0.5 $M_\odot$ and the expected p-factors between 1.2 and 1.4. In our method both the mass and the p-factor are directly scaled with the assumed $A$. The use of scaling $p \sim A$ to estimate the mass of the Cepheid is described in Section~\ref{sub:mass}.

The components are distant enough from each other (at least 375 $R_\odot$) that  there are no proximity effects and the assumed separation has negligible influence on the values of the other fitted parameters.

The presence of significant third light was detected in all analyzed bands. The highest value, of about $5\%$, was obtained for MACHO $R_C$, but up to $10\%$ values are possible within the errors. The uncertainty of the third light determination significantly influences the uncertainties of the other parameters, especially the inclination and the sum of the radii.

To obtain an optimal solution and an error estimation a standard Monte Carlo and Markov Chain Monte Carlo (MCMC) sampling was used. Best model fits to different light curves are presented in Figures \ref{fig:oglei_mod} and \ref{fig:machor_ecl}.

\begin{deluxetable}{lcccc}
\tablecaption{Photometric parameters of OGLE-LMC-T2CEP-098 from the Monte Carlo simulations.}
\tablewidth{0pt}
\tablehead{
\colhead{Parameter} & \colhead{Mean(S1)} & \colhead{Solution 1} & \colhead{Mean(S2)} & \colhead{Solution 2}
}
\startdata
$P_{orb} (days)$&  -        &  397.178(3)    &  -         & {\bf 397.1781} \\
$T_{I} (days)$  &  -        & 1858.959(7)    &  -         & {\bf 1858.9591} \\
$i$ ($^\circ$)  &  -        &  87.15(5)      &  -         &   87.25(6)    \\
$r_1+r_2$       &  0.11235  & 0.1134(3)$^a$  &  0.1113    & 0.1123(4)$^a$ \\
$r_2/r_1$       &  1.043    & 1.022(16)$^a$  &  1.08      & 1.06(4)$^a$ \\
$j_{21}(I_C)$   &  3.79     & 3.11(3)$^a$    &  3.81      &  3.07(4)$^a$  \\
$j_{21}(R_C)_M$ &  5.41     & 4.28(11)$^a$   &  -         &  -            \\
$j_{21}(V)_M$   &  8.48     & 6.12((15)$^a$  &  -         &  -          \\
$e$             &  -        &  0.00(1)       &  -         &  {\bf 0.0}    \\
$\omega$        &  -        &  -             &  -         &  -          \\
$p$-factor      &  -        &  1.240(12)$^b$ &  -         &  1.222(15)$^b$ \\
$l_3 (I_C)$     &  0.026    &  0.025(15)$^a$ &  0.026     &  {\bf 0.025}$^{a}$ \\
$l_3 (R_C)_M$   &  0.053    &  0.051(17)$^a$ &  -         &  -          \\
$l_3 (V)_M$     &  0.018    &  0.017(17)$^a$ &  -         &  -          \\
\multicolumn{5}{l}{Derived quantities:}\\
$r_1$           &  0.0550   & 0.0561(5)$^a$  &  0.0534    & 0.0544(10)  \\
$r_2$           &  -        & 0.0573(5)$^a$  &  -         & 0.0579(10)  \\
$L_{21}(I_C)$   &  4.1352   &  3.25(11)$^a$  &  4.52      & 3.49(23)      \\ 
$L_{21}(R_C)$   &  5.8847   &  4.47(18)$^a$  &  -         &  -   \\ 
$L_{21}(V)$     &  9.1971   &  6.40(23)$^a$  &  -         &  -  
\label{tab:photpar}
\enddata
\tablecomments{Epoch of the primary eclipse $T_{I}$ is $HJD - 2450000$~d, $L_{21}$ is the light ratio of the components in every photometric band. Fixed values are given in bold font. \added{Component 1 is the Cepheid.} }
\tablenotetext{a}{ values correspond to a pulsation phase 0.0}
\tablenotetext{b}{ for an assumed Cepheid mass of 1 $M_\odot$ and separation of 437 $R_\odot$}
\end{deluxetable}

\subsection{Results} \label{sub:results}

In Table~\ref{tab:photpar} we present two solutions. The first (Solution 1 or S1) was obtained using OGLE $I_{C}$ together with MACHO red and blue band data transformed to $R_C$ and $V_J$, respectively. For the second one (Solution 2 or S2) we used only the OGLE $I_{C}$ data with period, $T_I$, and the third light fixed as the former ones are band-independent and benefit from longer baseline and the latter is hard to establish using only one-band. For each solution both fitted and mean values are given for the parameters that vary during the pulsation cycle. The fitted parameters describe the system at  pulsation phase 0.0 using the following ephemeris:

  $$T_{max} (HJD) = 2455500.000 + 4.973726\times E$$.

Solution 2 serves as a comparison and to estimate possible systematic errors resulting from the inclusion of the non-standard MACHO photometry.  Note that not including all the parameters in the fitting procedure makes the estimated errors for the second solution much lower than they really are. 
The results are quite consistent with only small differences between the majority of parameters. The largest discrepancy was found for the mean sum of the radii -- about $2\sigma$ of the combined uncertainties.
From the photometric solution the relative radius change can be calculated to be about 25\% of the mean radius. This value is independent of the assumed $A$, as it comes directly from the modeling of the light curves and the pulsational radial velocity curve. We emphasize that different separations lead to different p-factor values.

\subsection{Colors and temperatures}
\label{sub:temp}

The system and individual magnitudes were dereddened using a value of the ratio of total to selective extinction of $R_V =  A_V / E(B-V) = 3.1$ and assuming an average combined LMC and Galactic foreground $E(B-V)$ value of 0.12 mag \citep{2007ApJ...662..969I}. We used the interstellar extinction law relations provided by \cite{1989ApJ...345..245C}, i.e.: $A_R/A_V = 0.75$, $A_I/A_V = 0.48$ and $A_K/A_V = 0.114$. The results are presented in Table~\ref{tab:phot}.

In case of OGLE $V$-band data we obtained the light ratio by using the S1 model, fixing the surface brightness ratio at the value obtained for the MACHO $V$-band data and fitting just the third light. The best fit was found for no third light at all. Apart from the $(V-R)$ color in all cases where 
$V$-band magnitude was used it was an average of OGLE and MACHO, i.e. 16.82 for the Cepheid and 14.44 for the companion.

Following \cite{macho2002alcock} we used two semi-empirical transformations from \cite{1993ApJS...86..541C} to obtain temperatures using $(V-I)$ and $(V-R)$ colors from OGLE and MACHO projects, respectively:
 
$$ \log(T_{eff}) = 4.199 - \sqrt{0.04622 + 0.2222(V-I)} $$

$$ \log(T_{eff}) = 4.199 - \sqrt{0.08369 + 0.3493(V-R)} $$

Temperatures calculated from both colors gave the same temperature for the Cepheid ($\sim 5300$ K) and values of $\sim 9000$ and $\sim 10400$ K for the companion. These results are consistent with the temperatures obtained from the limb darkening fitting.
We have also tested different calibrations from \cite{2009A&A...497..497G} and \cite{2005ApJ...626..465R}, and obtained for the Cepheid temperature values between 5200 and 5300 K.

For the VMC $K$-band data we used the known temperature and colors to estimate the $(V-K)$ color of the Cepheid, for which the results are more consistent. Using this color we calculated the K-band magnitude of the Cepheid, and then subtracting its light from the dereddened system light we could calculate the brightness of the blue companion. Two values are given, one without the third light and the second (in parenthesis) for an assumed $10\%$ of system light contributed by the third light. The assumed value would include both the unseen third companion and a possible Cepheid envelope,  known to produce an infrared excess of that order \citep{2012A&A...538A..24G}. The temperatures obtained for both of these cases are also given.

\begin{deluxetable}{lccccc}
\tablecaption{Photometric properties and temperatures of the components.}
\tablewidth{0pt}
\tablehead{
 \colhead{Parameter} & \colhead{System}& \colhead{Cepheid}& \colhead{Companion}& \colhead{$3^{rd}$ light}& \colhead{Unit}}
\startdata
\multicolumn{6}{l}{OGLE:}\\
 $I_{C}$          &  14.37          & 16.17            & 14.65              & 18.3               & $mag$     \\
 $V$              &  14.67          & 17.18            & 14.78              &   -                & $mag$     \\
 $I_{C} - A_I$    &  14.19          & 15.99            & 14.47              & 18.2               & $mag$     \\
 $V - A_V$        &  14.30          & 16.80            & 14.41              &   -                & $mag$     \\
 $(V-I)_{ogle}$   &   0.11          &  0.81            & -0.061             &   -                & $mag$     \\
\multicolumn{6}{l}{MACHO:}\\
 $R_{C}$         &  14.59          & 16.71            & 14.81              & 17.8               & $mag$     \\
 $V$             &  14.71          & 17.21            & 14.84              & 19.1               & $mag$     \\
 $R_{C} - A_R$   &  14.31          & 16.43            & 14.54              & 17.5               & $mag$     \\
 $V - A_V$       &  14.34          & 16.84            & 14.47              & 18.7               & $mag$     \\
 $(V-R)_{macho}$ &   0.03          &  0.41            & -0.064             &  1.22              & $mag$     \\
\multicolumn{6}{l}{VMC:}\\
 $K$                &  13.89          &   -              &   -                &   -                & $mag$     \\
 $K - A_K$          &  13.85 (13.96)  & 15.02$^a$        & 14.30 (14.49)      & (16.35)            & $mag$     \\
 $(V-K)_{vmc}$      &   0.47 (0.36)   &  1.8             &  0.14 (-0.05)      &   -                & $mag$     \\
\multicolumn{6}{l}{Temperatures:}\\
 $T(V-I)_{ogle}$    &  8600           &  5300            & 10400              &                     & $K$     \\
 $T(V-R)_{macho}$   &  7800           &  5300            &  9000              &                     & $K$     \\
 $T(V-K)_{vmc}$     &  7900 (8300)    &  5300$^a$        &  9000 (10400)      &                     & $K$     \\
 $T_{LD}$$^b$       &                 &  5550            &  9500              &                     & $K$     
\label{tab:phot}
\enddata
\tablecomments{ System, Cepheid and companion observed and dereddened magnitudes are given. For the Cepheid magnitudes, colors and temperatures are mean values over the pulsation cycle.}
\tablenotetext{a}{ fixed or calculated directly using a fixed value}
\tablenotetext{b}{ \added{temperature obtained from the limb darkening analysis}}
\end{deluxetable}


\begin{figure}
\begin{center}
  \resizebox{0.6\linewidth}{!}{\includegraphics{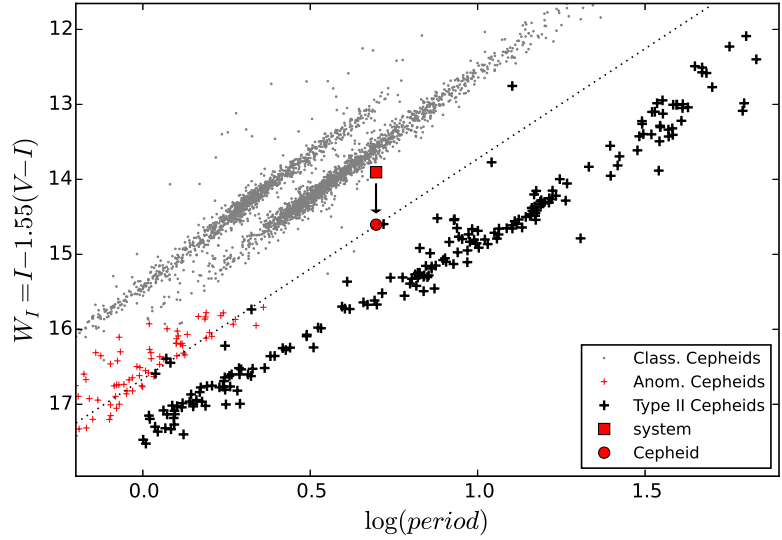}} \\
\caption{Position of the system and the Cepheid on the period-luminosity diagram for Cepheids in the LMC. Even after the subtraction of the companion light the Cepheid is far from the PL relation for type II Cepheids, located almost exactly between that relation and the one for classical Cepheids. Its position resembles that of Anomalous Cepheids. An assumed P-L relation for 1.5 $M_\odot$ pulsators is marked with the dotted line.
\label{fig:perlum}}
\end{center}
\end{figure}

Once the light of the stars composing the system were disentangled it was possible to check how the position of the Cepheid in the period-luminosity diagram has changed. As seen in Fig.~\ref{fig:perlum} the star moves closer to the P-L relation for type II Cepheids, it is still too far away to be consistent with that relation.
It is interesting to note that the application of a similar correction (subtraction of the companion light) would move all but one other outliers onto the P-L relation for type II Cepheids. 

\added{
We have also checked the position of the Cepheid on the K-band P-L diagram published by \citet{2017AJ....153..154B} -- see their Fig.9. After the subtraction of the companion light the Cepheid lies closer to the relation for type II Cepheids than to the one for the classical Cepheids, but is still one of the most outlying points. In this case however even half of the same correction as applied for T2CEP-098  would move all the other outliers onto their P-L relation.}

It is not very likely that the system is located closer than the LMC, as the systemic velocity of $\sim 260$ km/s corresponds perfectly to the average LMC velocity. One of the possibilities could be that in case of T2CEP-098 some anomalous, extreme reddening may be present, but other explanations, including a different evolutionary state, cannot be excluded. In fact, the position of the Cepheid between the period-luminosity relations of classical and type II Cepheids resembles the position of Anomalous Cepheids, which are interpreted as mergers with masses about 1.5 $M_\odot$. Such a value for the mass corresponds perfectly to our best estimate for the mass of the Cepheid, which is discussed in the next section.

\begin{figure}
\begin{center}
  \resizebox{0.4\linewidth}{!}{\includegraphics{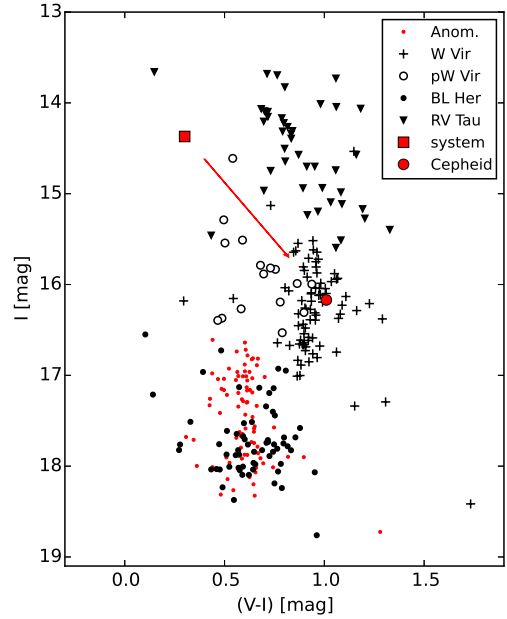}} \\
\caption{Position of the system and the Cepheid on the color-magnitude diagram for type II Cepheids in the LMC. After the subtraction of the component light the Cepheid moves to the zone occupied by typical W Virginis stars. It is possible that all the outliers on this plot are members of binary systems with bright companions of very different color.
\label{fig:cmd}}
\end{center}
\end{figure}

In the color-magnitude diagram (Fig.~\ref{fig:cmd}) the Cepheid in the system (with the companion) lies to the blue of the overall distribution of type II Cepheids. Correction for the light of the companion, however, moves the Cepheid to the region occupied by the normal W Virginis stars.
This means that the color alone cannot serve to discriminate the peculiar variables (as suggested by \citealt{2008AcA....58..293S}). Discrepant color might suggest that the star may be a member of a binary system. It is even possible that the majority (if not all) of the outliers on this plot are members of binary systems with a quite bright companion of very different color. 

\pagebreak

\section{Discussion} \label{sec:summary}

\subsection{Mass estimation}
\label{sub:mass}

The bright early-type companion is only seen in the Balmer lines. Its detection is hindered by the presence of about 3-4 other sources seen in the spectra. One of these has the radial velocity in the range of the most probable quadrature velocities of the companion making it even harder to detect \added{(see Fig.~\ref{fig:rvorb}). Velocities measured from the Balmer lines were not accurate enough to enable us to measure an orbit for the companion.} Spectra with very high signal to noise ratios would be required together with careful spectral disentangling to try to identify the Cepheid companion and measure its velocity.

In our method, however, we relate the absolute radius change $\Delta R(t,p)$ (see Fig.~\ref{fig:rvpuls}) obtained from the integration of the pulsational radial velocity curve (depending on the p-factor) to the relative change of the radius $\Delta r(t)$ obtained from the photometry, by the star separation $A$:

$$ \Delta R(t,p) = \Delta R(t,p=1) * p = A * \Delta r(t) $$.

As the shape of  $\Delta r(t)$ is taken from integration of the radial velocity curve, what actually matters are just the amplitudes of the radius change, i.e. $\Delta R(p=1) = 4.83 R_\odot$ and $\Delta r = 0.01375$.

If we knew the separation of the stars in the system then from this analysis we could obtain the value of the p-factor. But we can invert this equation: assuming a value for the p-factor we can obtain the separation of the stars in the system:

\begin{equation} 
 A = p * \frac{\Delta R(p=1)}{\Delta r} = p * 351.987, 
\label{eqn:ap}
\end{equation}

\noindent
which lets us directly calculate the masses and radii of the Cepheid and its companion:

$$ \mean{R}_{cep} = \mean{r}_{cep} * A = 0.055 * A = 19.36 * p $$
\added{$$ \mean{R}_{2} = \mean{r}_{2} * A = 0.0573 * A = 20.17 * p $$}
$$ M_{cep} = 3.705*(p^3 - 1.065*p^2) $$
\added{$$ M_{2} = 3.946*p^2 $$}

\noindent
The equations for the masses are derived from the orbital solution (with the known inclination) and Kepler's third law substituting the dependence on the semi-major axis with the one on the p-factor (equation~\ref{eqn:ap}).

\begin{figure}
\begin{center}
  \resizebox{0.48\linewidth}{!}{\includegraphics{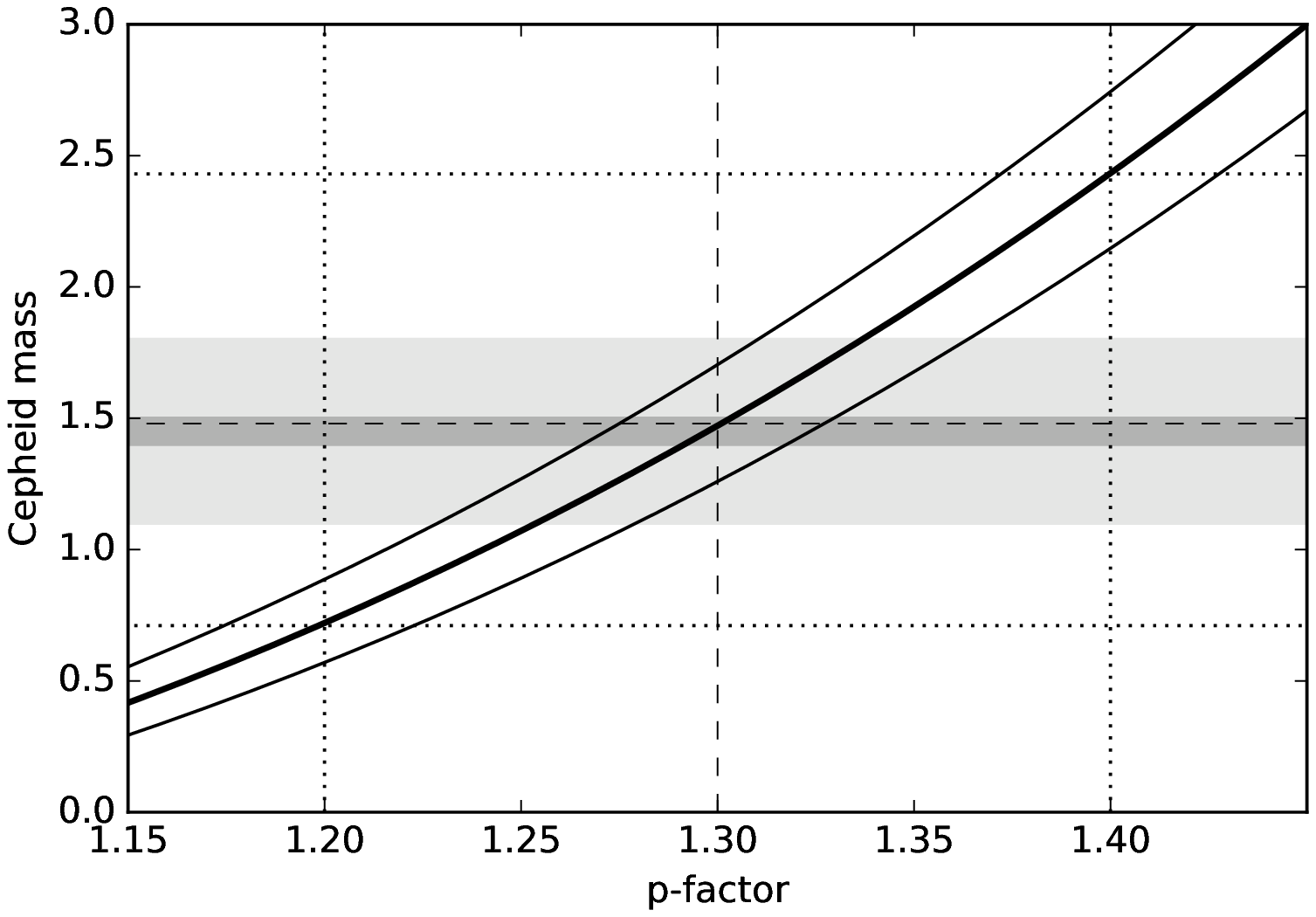}} 
  \resizebox{0.48\linewidth}{!}{\includegraphics{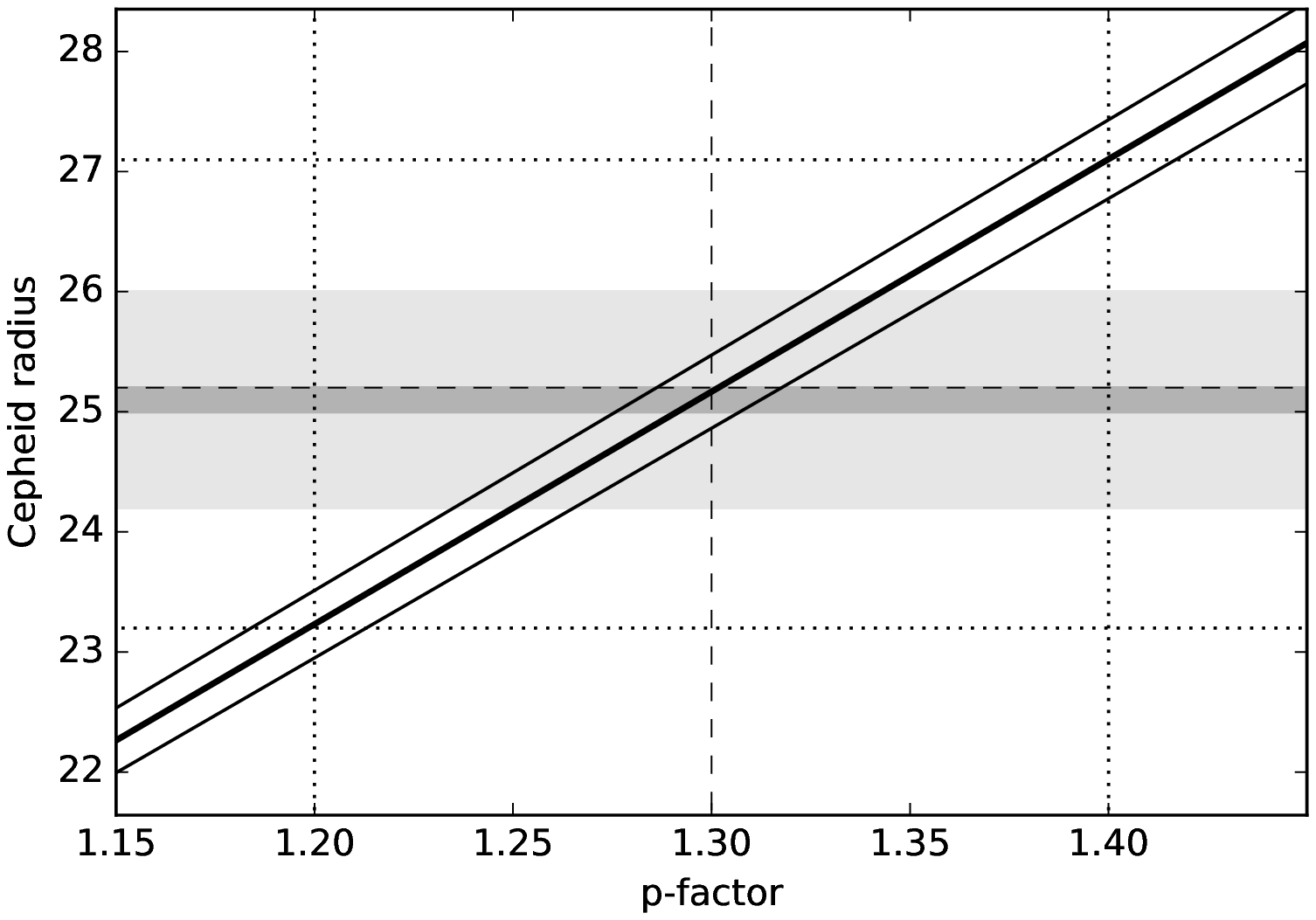}} \\  
\caption{ The p-factor versus radius ({\em left}) and mass ({\em right}) relations. In our method we can relate the p-factor value with the stars separation and thus with other physical parameters. The values of $p$ lower than 1.2 and higher than 1.4 can be safely excluded, which limits the mass to the range between 0.72 and 2.43 $M_\odot$. Dark gray areas marks the combined best solutions from the bolometric correction and SBCR methods (see text), and the light gray areas marked their combined errors.
\label{fig:pfac_mass}}
\end{center}
\end{figure}

We can put limits on these relations. First, the separation cannot be smaller than the orbit radius of the Cepheid, which is $375 \pm 3 R_{\odot}$. Second, the p-factor itself has  limits. Taking into account our studies of classical Cepheids in binary systems we can assume that the value of the p-factor falls between 1.2 and 1.4.
This assumption is in agreement with other recent measurements of p-factors \citep{2016A&A...587A.117B, 2017A&A...597A..73N, 2017arXiv170105192K}. Using all high-accuracy determinations from the literature \citet{2017arXiv170105192K} calculated a weighted average for this parameter to be $p = 1.29 \pm 0.04$, while the previous period-projection factor relation derived by \citet{2009A&A...502..951N} gives $p = 1.25 \pm 0.05$.

Thus we can estimate the minimum and maximum mass of the Cepheid and its companion as well as limits for other physical parameters. Parameters for three cases (for p-factors 1.2, 1.3 and 1.4) are presented in Table~\ref{tab:abs} and the graphical form of the p-factor~-~mass and the p-factor~-~radius dependence is shown in Fig.~\ref{fig:pfac_mass}. Note that the estimated uncertainties of parameters derived from the orbital solution include the p-factor uncertainty through the $A \sim p$ scaling and therefore are highly correlated, i.e. if the mass of the Cepheid were lower, the mass of the companion, the semi-major axis and the radii would be lower as well.

\begin{deluxetable}{lcccc}
\tablecaption{Physical parameters of OGLE-LMC-T2CEP-098.}
\tablewidth{0pt}
\tablehead{
 \colhead{Parameter} & \colhead{p = 1.2} & \colhead{p = 1.3} & \colhead{p = 1.4} & \colhead{Unit}}
\startdata
\multicolumn{5}{c}{Cepheid} \\
mass             & 0.72(16)     & 1.47(21)  & 2.43(27)  & $M_\odot$ \\ 
radius           & 23.2(4)      & 25.2(4)   & 27.1(4)   & $R_\odot$ \\ 
$\log g$         & 1.56(10)     & 1.80(7)   & 1.96(5)   & cgs       \\ 
luminosity       & 0.38(3)    & 0.45(4)  & 0.52(4) & $10^3 L_\odot$ \\ 
$M_{bol}$        & -1.70(9)     & -1.88(9)  & -2.04(9)  & mag       \\
$BC_V$ (calc.)   & -0.02(10)    & -0.20(10) & -0.36(10) & mag       \\ 
temperature      & \multicolumn{3}{c}{$5300 \pm 100$}   & K         \\
SpType           & \multicolumn{3}{c}{ G9 III}          &           \\
$M_V$              & \multicolumn{3}{c}{-1.68}          & $mag$     \\
\multicolumn{5}{c}{Early-type companion} \\
mass             & 5.7(3)       & 6.7(4)    & 7.8(4)    & $M_\odot$ \\ 
radius           & 24.2(3)      & 26.2(4)   & 28.2(4)   & $R_\odot$ \\ 
$\log g$         & 2.42(3)      & 2.42(3)   & 2.42(3)   & cgs       \\ 
luminosity       & 4.3(0.9)     & 5.0(1.1)   & 5.8(1.2) & $10^3 L_\odot$ \\
$M_{bol}$        & -4.33(23)    & -4.50(24) & -4.66(22) & mag       \\
$BC_V$ (calc.)   & -0.27(23)    & -0.44(24) & -0.60(22) & mag      \\ 
temperature      & \multicolumn{3}{c}{$9500 \pm 500$}   & K     \\
SpType           & \multicolumn{3}{c}{ B9/A0 III}        &      \\
$M_V$              & \multicolumn{3}{c}{-4.06}           & $mag$     \\
\multicolumn{5}{c}{System} \\
semimajor axis   & 422(7)       & 458(7)   & 493(7)    & $R_\odot$ \\ 
orbital period   & \multicolumn{3}{c}{397.178(3)}      & $days$ 
\label{tab:abs}
\enddata
\tablecomments{ \added{Three cases for different p-factors are presented.} For the Cepheid the radius, gravity ($\log g$), magnitudes, and colors are mean values over the pulsation cycle. The errors include the p-factor uncertainty.}
\end{deluxetable}

At this point our best estimate for the Cepheid mass is about $1.4 M_\odot$. It is not the expected mass for type II Cepheids, but as already noted, the star is probably not a normal type II Cepheid. A mass of 0.5-0.6 $M_\odot$, however, is still a possible value for the Cepheid if the p-factor is closer to $1.2$.

We can also make another kind of analysis. Because we know the distance modulus to the LMC to high precision -- $18.49 \pm 0.01(stat) \pm 0.05(sys)$ mag \citep{2013Natur.495...76P}, we can calculate the luminosities of the stars and compare them with those calculated for different assumed p-factors (or star separations).

The system lies close to the galactic center (about 30 $arcsec$) and to the line of nodes of the LMC, so the geometric correction calculated using the \cite{2002AJ....124.2639V} model is small -- only 0.01 mag. Using a different model of the LMC \citep{2016ApJ...832..176I} we obtained a higher \added{angular} distance from the center ($\sim 55 arcsec$), but the same geometric correction. In both cases the system is located farther than the LMC center.
We adopt a distance modulus to the binary of $18.50 \pm 0.05$ mag.

We can use this value to calculate the absolute magnitudes ($M_V$) of the system components using the observed dereddened magnitudes from Table~\ref{tab:phot}. Using the luminosities calculated for different radii we can also calculate the bolometric magnitudes ($M_{bol}$) and derive the necessary bolometric correction:

$$ BC_V = M_{bol} - M_V $$.

These values are presented in Table~\ref{tab:abs}. Now we can compare the values of $BC_V$ with those expected for stars of these types. From the calibration of \cite{1999A&AS..140..261A}  we obtain a value of $-0.20 \pm 0.02$ for the Cepheid and  $-0.18 \pm 0.05$ for the companion.
The $BC_V$ for the Cepheid strongly favors the central solution for $p=1.3$ presented in Table~\ref{tab:abs} and implies a mass of the Cepheid of $1.5 \pm 0.4 M_\odot$ and  $p = 1.3 \pm 0.05$.  Because of the large errors  conclusions regarding  $BC_V$ for the companion are not so definitive. 

Knowing the distance we can also calculate the star radii using the surface brightness - (V-K) color relation (hereafter SBCR). Using the formulas from \cite{2014A&A...570A.104C} and \cite{2005MNRAS.357..174D}, for a Cepheid color $(V-K) = 1.8 \pm 0.09$ and a distance 50.1 kpc we obtain a radius of $R_{cep} = 25.0 \pm 1.2 R_\odot$, which translates to $M_{cep} = 1.4 \pm 0.5 M_\odot$ and $p = 1.29 \pm 0.06$. The results are not very precise because of the low accuracy of the (V-K) color. \added{Note that the flux ratio in K-band was not directly measured and the K-band magnitude of the Cepheid had to be estimated.}

The (V-K) color for the companion is  poorly established, and the resulting radius from the SBCR is $R_{comp} = 24.9 \pm 2.5 R_\odot$.
Once again, the values obtained for the Cepheid strongly indicate the central solution, and those for the companion suggest a solution between the lower limit and the central solution. In Fig.~\ref{fig:pfac_mass} the dark gray areas show the combined best solutions from the bolometric correction and the SBCR method, and the light gray areas show the possible parameters within the combined errors of these two methods.

\subsection{Parameters of T2CEP-098 constrained with stellar pulsation theory}
\label{sub:puls_theory}

We can further constrain the parameters of T2CEP-098 using stellar pulsation theory. T2CEP-098 pulsates in the radial fundamental mode and its pulsation period, $P\!\approx\!4.974$\thinspace d, is actually the most accurate observable at hand. Pulsation periods can be precisely reproduced with linear pulsation codes as is commonly done in asteroseismic studies of multimode pulsators of various types \citep[see e.g.][]{asterobook}, including classical pulsators \citep{md05}. Below we check whether each $p$-factor value in the $1.2-1.4$ range, considered in Sect.~\ref{sub:mass}, or alternatively whether each mass value in the $0.72-2.43\MS$ range, leads to a pulsation period (computed with the linear pulsation code) consistent with the observed period. Uncertainties intrinsic to the modeling (e.g. non-linear period shifts, turbulent convection model) will be discussed later on and taken into account during comparison of the computed and observed periods.

For the computation of pulsation periods we use the linear non-adiabatic pulsation code of \cite{sm08a} which implements the time-dependent, turbulent convection model of \cite{kuhfuss}.  This code, together with its non-linear version, has been extensively used to study the dynamical behavior of type-II Cepheids -- see \cite{s16} and references therein. Here we use exactly the same model parameters (structure of numerical grid and parameters of the turbulent convection model) as in the above study. All parameters are explicitly given in section 3 of  \cite{blherPD}, in particular the convective parameters are those of set P1. In all model computations we use the OPAL opacities \citep{opal} supplemented at low temperatures with the opacity data from \cite{ferguson} and computed for the scaled solar chemical composition as given by \cite{a09}. To compute the pulsation period, the necessary input parameters are mass, $M$, luminosity (or radius), $L$ (or $R$), effective temperature, $T_{\rm eff}$, and chemical composition (hydrogen and metal abundance, $X$ and $Z$). 

Before we proceed to the modeling of T2CEP-098, we first demonstrate how the pulsation code in this
approach reproduces the pulsation period of OGLE-LMC-CEP-0227, a fundamental mode classical Cepheid and a member of an SB2 eclipsing binary system. \cite{cep227mnras2013} derive $M=4.165\pm0.032\MS$, $T_{\rm eff}=6050\pm 160$\thinspace K and $\log L/\LS=3.158\pm0.049$. The metallicity of this system is unknown and so we adopt ${\rm [Fe/H]}=-1.0$. With these parameters, the computed fundamental mode period is $3.820$\thinspace d compared with the observed period of $3.797$\thinspace d. By increasing the effective temperature by just $10$\thinspace K, the pulsation calculations yield $3.798$\thinspace d. For classical Cepheids non-linear period shifts are small, of order of a fraction of a percent. The agreement between the observed and computed periods is thus excellent.

For T2CEP-098 the mass and radius are functions of the $p$-factor (see Fig.~\ref{fig:pfac_mass}), while its effective temperature was estimated in Sect.~\ref{sub:temp} to be $T_{\rm eff}=5300\pm 100$\thinspace K. We have no metallicity estimate for T2CEP-098 and hence we  consider three different values, ${\rm [Fe/H]}=0.0, -1.0$ and $-2.0$, covering both the metal rich and metal poor scenarios.

\begin{figure}
\centering
\includegraphics[width=10cm]{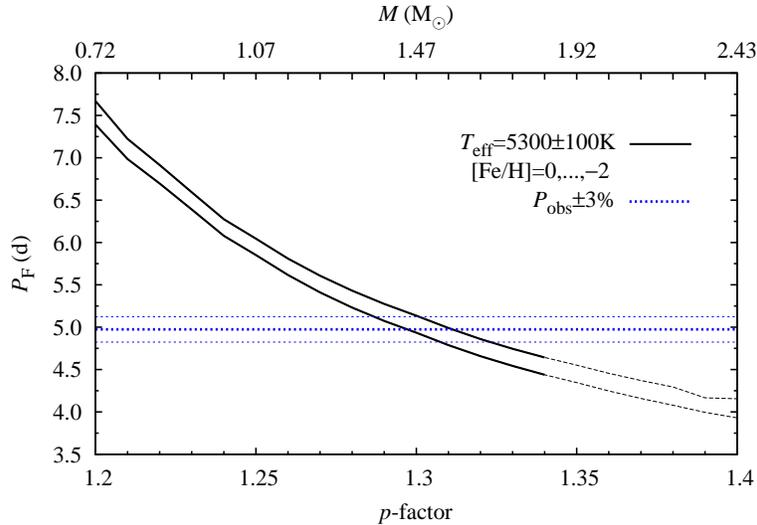}
\caption{The ranges of fundamental mode period allowed by stellar pulsation calculations plotted versus $p$-factor. At each $p$-factor, the minimum and maximum fundamental mode period in a set of models with different effective temperatures and different metallicities is plotted. Thick solid lines correspond to models within the instability strip, while thin dashed lines correspond to models beyond the red edge. Model predictions are compared with the pulsation period observed for T2CEP-098 (thick and dotted horizontal line). The thin horizontal lines reflect the uncertainty inherent to pulsation modeling.}
\label{fig:permod}
\end{figure}

For each value of $p$-factor (and hence for each corresponding value of $M$/$R$) computed in the $1.2-1.4$ range  (with a $0.01$ step), we have computed a set of pulsation models with effective temperatures of $5200$, $5300$ and $5400$\thinspace K. For each effective temperature, three metallicities (${\rm [Fe/H]}=0.0, -1.0$ and $-2.0$) were adopted, yielding nine models at a given value of $p$-factor. The resulting minimum and maximum fundamental mode periods for these models are plotted as a function of $p$-factor in Fig.~\ref{fig:permod}. For convenience, the corresponding mass values are plotted along the top axis. The solid line segments correspond to $p$-factors for which at least one model in a set is linearly unstable, i.e. is located within the instability strip. The dashed line segments correspond to $p$-factors for which all models are linearly stable, i.e. are located on the red side of the instability strip. Models with periods close to that of T2CEP-098 are located within the instability strip, close to its red edge.

Before the range of the fundamental mode periods calculated by the pulsation models at a given value of $p$-factor/mass is compared to the observed period, one needs to consider the accuracy of the computed linear periods. We are using a pulsation code with a relatively coarse Lagrangian numerical grid, made necessary by the time-consuming, non-linear computations. In particular, we use only 150 zones, of which the outer 40 have equal mass, down to the anchor zone in which the temperature is fixed to $11000$\thinspace K and then 110 zones with mass geometrically increasing inward, down to $2\!\times\!10^6$\thinspace K. If these mesh parameters are altered, e.g. by increasing the number of zones or extending the envelope down to $4\!\times\!10^6$\thinspace K, etc., then the period of the fundamental mode changes by no more than $0.5$\thinspace per cent. Another uncertainty arises from the simple one equation convection model we use, which contains several free parameters. Only the overall efficiency of the convection, determined by the mixing-length parameter, $\alpha$, can alter the pulsation periods significantly. By changing the default value we adopt in the computations ($\alpha=1.5$) by $\pm0.3$, the period of the fundamental mode changes by up to $\pm2.5$ per cent. Finally, we compute  linear periods, while the observed period corresponds to full-amplitude, non-linear oscillations. The expected non-linear period shifts for type-II Cepheids were studied in detail by \cite{s16}, (see in particular their Fig.~6). Non-linear period shifts can be as large as $\pm 15$\thinspace per cent, but only for more luminous, longer-period stars. For T2CEP-098, which is  located close to the red edge of the instability strip  with a pulsation period close to $5$\thinspace d, we may expect that the non-linear period should be slightly shorter, by up to $2$\thinspace per cent.

To take into account the above uncertainties related to stellar pulsation calculations, we require that the computed periods match the observed period to within $\pm3$\thinspace per cent, which is marked with thin horizontal lines in Fig.~\ref{fig:permod}. It is clear that stellar pulsation calculations strongly limit the range of possible values of $p$-factor and consequently of mass, even though the metallicity of the Cepheid is unknown. The $p$-factor must be in the range $1.287-1.323$, and consequently the mass in the range $1.45-1.57\MS$, for the pulsation models to agree with the observed period. Adding the uncertainty of the p-factor from Section \ref{sub:results} and the uncertainty of radius and orbital elements that influence the mass-radius relation, we estimate the mass of the Cepheid to be $1.51 \pm 0.09 M_\odot$ and $p = 1.30 \pm 0.03$.

We have shown that by using linear pulsation calculations we can strongly constrain the parameters of this single-lined eclipsing binary system. Stellar pulsation theory offers more possibilities. Non-linear calculations of the radial pulsation are possible, which may lead to further refinement of the parameters of T2CEP-098 through comparison of the observed and the modeled light and radial velocity curves. Such modeling is time-consuming and is beyond the scope of the present analysis. A dedicated study of this and other similar systems is planned.

\subsection{Classification} \label{sub:class}

The classification of the Cepheid T2CEP-098 is complicated as it shows characteristics of various classes of pulsators. The estimates of physical parameters presented above indicate a mass of 1.4-1.5 $M_\odot$, as expected for Anomalous Cepheids \citep{1997AJ....113.2209B}. Classical Cepheids have masses closer to 4 $M_\odot$, and type II Cepheids are expected to have masses around 0.5 $M_\odot$.
Another indication that the star might be a long-period Anomalous Cepheid is its position on the period-luminosity diagram. As seen in Fig.~\ref{fig:perlum} the star is located exactly between the classical and type II Cepheids, similar to Anomalous Cepheids. A common period-luminosity relation can be even determined for these 1.5 $M_\odot$ pulsators, parallel to the PL relations for the other types.

On the other hand the color of the star is quite different. Anomalous Cepheids occupy a narrow color range around $(V-I) \sim 0.6$, while the corrected color for the pulsating component is about 1.0 (reddened) -- see Fig.~\ref{fig:cmd}. As already stated in subsection~\ref{sub:photo} the shape of the light curve of T2CEP-098 is more similar to BL Herculis (or even classical Cepheid) type stars than to W Virginis stars. There are no known Anomalous Cepheids with such long periods, so the shapes of the light curves cannot be compared.

\begin{figure}
\begin{center}
  \resizebox{0.6\linewidth}{!}{\includegraphics{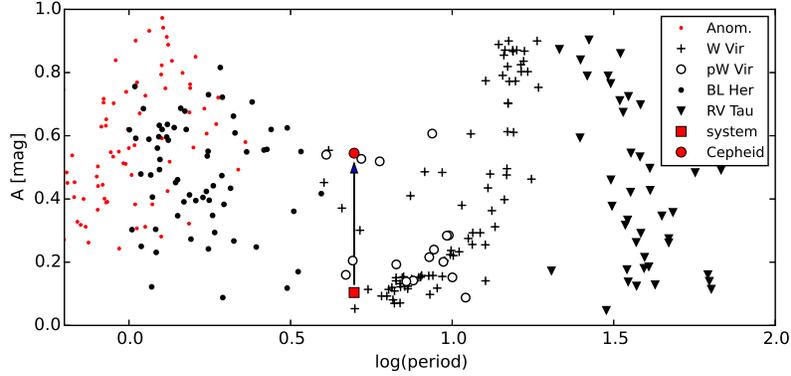}} \\
\caption{Period-amplitude diagram for Type II Cepheids. The apparent amplitude has changed significantly after subtraction of the companion light suggesting that the star has more in common with BL Her stars, or even Anomalous Cepheids, which on average have amplitudes about 0.5-0.6 mag.
\label{fig:four_amp}}
\end{center}
\end{figure}

The difference between T2CEP-098 and the W Virginis stars can be also seen when we correct the amplitude of pulsations by subtracting the light of the bright blue companion (see Fig.~\ref{fig:four_amp}). The star moves from a position typical for W Vir stars with periods of about 5 days (amplitude $\sim$ 0.1 mag) to a position typical for BL Her stars or Anomalous Cepheids (amplitude $\sim$ 0.5-0.6 mag on average). This suggests that its classification as a peculiar W Virginis is uncertain.

A similar situation may be also true for other stars tagged by \cite{2008AcA....58..293S} as WVir or pWVir based only by a period criterion. More detailed studies and analysis of other eclipsing type II Cepheids will be necessary to answer this question.

The problem with the classification of T2CEP-098 may be resolved if we consider an evolutionary history  different than for standard classical or type II Cepheids. The first hint for such a solution is its similarity with Anomalous Cepheids, which are thought to originate in a merger of two lower-mass stars. Another hint is its very unusual configuration, resembling the Algol paradox, where the less massive (1.5 $M_\odot$) star is more evolved than the more massive star (6.8 $M_\odot$). The lower-mass Cepheid seems to have passed through the red giant stage, while the more massive component apparently has just left the Main Sequence and is now moving redward through the Hertzsprung gap.

To explain the current state of the system, mass transfer seems to be necessary. One  possible scenario is that the originally more massive ($\sim 5 M_\odot$) Cepheid progenitor filled its Roche lobe on its way up the red giant branch  (at an age of about 100 Myr) and started transferring mass to its companion. At this stage the orbital period would be about 100 days and the separation between the stars about 180 $R_\odot$. The stars would pass through  mass reversal (which was also necessary to resolve the Algol paradox) and the current Cepheid would start burning helium and shrinking after transferring most of its mass to the companion, eventually passing through the instability strip. Its companion would grow from about 3.2 $M_\odot$ to the current 6.8 $M_\odot$ and after burning all the hydrogen in its core would leave the Main Sequence. This explanation assumes conservative mass transfer, and would need modification if a significant amount of mass and angular momentum were lost from the system.

This binary evolution scenario  would solve most of the aforementioned classification problems. To better constrain the evolutionary state of this system however, better observational data is necessary together with  more complex modeling. High quality spectra would allow a determination of the metallicity and more advanced pulsation models (utilizing non-linear calculations) would allow a better estimate of the physical parameters of both stars. K-band photometry covering the eclipses would greatly help in determining the temperatures of both components.

We make one general comment regarding the binarity of type II Cepheids. Fig.~\ref{fig:four_amp} shows a large scatter of the amplitudes of BL Her stars with many outliers with low amplitudes and with a clear upper limit at about 0.7 mag. The increase of the T2CEP-098 pulsational amplitude after the subtraction of the companion light and typically bluer colors of BL Her stars may suggest that some of the lower-amplitude objects may be members of binary systems. For example all of the BL Her stars outlying upwards on the period-luminosity diagram (see Fig.~\ref{fig:perlum}) have amplitudes lower than 0.35 mag.

\section{Summary and conclusions} \label{sec:conclusions}

We have presented an analysis of T2CEP-098, a binary system composed of a type II Cepheid, previously classified  as a peculiar W Virginis star, and a blue giant.
Analysis of spectroscopic data and the use of pulsation theory has allowed an improvement in our understanding of this system compared to the study made by \citet{macho2002alcock}. More and higher quality photometric data further improved the quality of the models.

The companion to the Cepheid is an early-type star of spectral type A0. For such stars metal lines are hardly present, but the hydrogen Balmer lines are very strong. We could detect the presence of the companion in these lines, but an accurate radial velocity curve for this star could not be measured due to its very low orbital amplitude and contamination from additional light sources in the spectra. Higher signal to noise ratio spectra are necessary to allow an analysis of the system as an SB2 binary.

Although the system is a single-lined spectroscopic binary we could determine  the most important physical parameters, including the masses, using a variety of independent methods. The mass of the pulsating component was estimated to be $1.51 \pm 0.09 M_\odot$, with pulsation theory yielding the strongest constraints. The corresponding mass of the companion was  calculated to be $6.8 \pm 0.4 M_\odot$.

In our method the physical parameters scale with the p-factor, which means that determining the mass of the Cepheid allows an estimation  of this important parameter. Our best estimate is $p = 1.30 \pm 0.03$, which is consistent with the recently measured value of $1.26 \pm 0.07$ for the type II Cepheid $\kappa$ Pavonis \citep{2015A&A...576A..64B}. This comparison must be tempered with the unclear classification of T2CEP-098.

The Cepheid has several features that make it similar to type II Cepheids. These include the amplitude and the shape of the light curve (similar to those of BL Her stars), possible erratic period change, a color typical of W Virginis stars, and a large relative radius change of $\sim 25\%$.
On the other hand, it lies exactly in the middle between the period-luminosity relations of classical and type II Cepheids, and has a mass close to 1.5 $M_\odot$, much larger than the mass of 0.5-0.6 $M_\odot$ expected for the type II Cepheids. The position on the P-L diagram and the mass may suggest that the star is a long-period Anomalous Cepheid.

The system configuration and especially the masses strongly suggest a binary interaction in the evolutionary history as  necessary to solve the Algol paradox.
Mass reversal after  mass transfer from the originally more massive component would explain the current evolutionary status of the components and solve all of the classification problems. A pulsating star that is a result of such a binary interaction might be called a Binary Evolution Pulsator, as suggested by \citet{2012Natur.484...75P} for a star with a similar history, or more specifically a Post-Algol Pulsator, as T2CEP-098 directly follows the evolution of an Algol-type system.

More and better quality data together with more advanced analysis is necessary for better understanding of the system and a more precise description of the pulsating component that although sharing some characteristics with classical, type II and Anomalous Cepheids, cannot be unambiguously classified as any of them.

Our study shows that in case of eclipsing binaries with pulsating components we can accurately derive the most important physical parameters of the stars even for single-lined binaries. This result is very important for the characterization of the unbiased sample of Cepheids and other pulsators.
Limiting analysis to double-lined systems places unnecessary restrictions to a limited number of  evolutionary scenarios and specific ranges of parameters. The inclusion of systems with only one component visible in the spectra will also make the sample of pulsating stars with measured physical parameters larger and statistically more significant. We have already identified several classical and type II Cepheids in this kind of systems, and plan appropriate followup studies.

\acknowledgments

We gratefully acknowledge financial support for this work from the Polish National Science Center grant SONATA 2014/15/D/ST9/02248. WG and GP acknowledge financial support for this work from the BASAL Centro de Astrofisica y Tecnologias Afines (CATA PFB-06/2007). WG also acknowledge support for this work from the Chilean Ministry of Economy, Development and Tourism's Millennium Science Initiative through grant IC120009 awarded to the Millennium Institute of Astrophysics MAS. PK acknowledges support from the French Agence Nationale de la Recherche (ANR), under grant ANR-15-CE31-0012-01. The research leading to these results has received funding from the European Research Council (ERC) under the European Union's Horizon 2020 research and innovation program (grant agreement No 695099).

This work is based on observations collected at the European Organisation for Astronomical Research in the Southern Hemisphere under ESO programmes: 096.D-0425(A), 097.D-0400(A) and 098.D-0263(A). We thank ESO and the CNTAC for generous allocation of observing time for this project.
We would also like to thank the support staff at the ESO Paranal and La Silla observatory and at the Las Campanas Observatory for their help in obtaining the observations.

This research has made use of NASA's Astrophysics Data System Service.

\vspace{5mm}
\facilities{ESO:3.6m (HARPS), VLT:Kueyen (UVES), Magellan:Clay (MIKE)}

\software{Python}

\appendix
\section{RaveSpan}

The analysis of the spectra and radial velocity curves presented above is based on the {\tt RaveSpan} code.

{\tt RaveSpan} is an easy to use graphical application that brings together three major velocity extraction methods: cross-correlation function (CCF), two-dimensional CCF (TODCOR) and Broadening Function (BF). All extracted velocities are instantly plotted in the RV curve window. Selected orbital parameters may be fitted afterwards.

{\tt RaveSpan} is composed of several components. There is a spectrum viewer, where one
can inspect collected spectra, compare them with templates, and choose a wavelength
range for the analysis. In the orbit viewer, one can see extracted velocities and a model
of the orbit. There is also an orbit fitting tool, which can be used to fit selected orbital parameters.
The radial velocity analysis tool allows users to see the output of CCF, TODCOR or
BF and interactively fit several profiles of different types. There is also a simple Spectral Disentangling mode implemented.

{\tt RaveSpan} is written in pure Python and uses PyQt graphical library with Matplotlib as a plotting tool. The code is available for download on a dedicated webpage\footnote{http://users.camk.edu.pl/pilecki/ravespan}, where more details are available.

\bibliographystyle{aasjournal}
\bibliography{t2ceppaper}

\listofchanges

\end{document}